\documentclass[journal=jacsat,manuscript=article]{achemso}

\usepackage{chemformula} 
\usepackage[T1]{fontenc} 
\usepackage[version=3]{mhchem}
\usepackage{epsf}
\usepackage{amsmath}
\usepackage{bm}
\usepackage{graphicx}
\usepackage{xcolor}

\newcommand{\onlinecite}[1]{\hspace{-2 ex} \nocite{#1}\citenum{#1}} 

\author{Zheng Sun}
\email{zsun@lps.ecnu.edu.cn}

\affiliation[University of Pittsburgh]
{Department of Physics and Astronomy, University of Pittsburgh, Pittsburgh, PA 15260, USA}
\alsoaffiliation[East China Normal University]
{State Key Laboratory of Precision Spectroscopy, East China Normal University, Shanghai, 200241, China}

\author{Jonathan Beaumariage}
\author{Qiaochu Wan}
\author{Hassan Alnatah}
\author{Nicholas Hougland}
\author{Jessica Chisholm}
\affiliation[University of Pittsburgh]
{Department of Physics and Astronomy, University of Pittsburgh, Pittsburgh, PA 15260, USA}

\author{Qingrui Cao}
\affiliation[Carnegie Mellon University]
{Department of Physics, Carnegie Mellon University, Pittsburgh, PA 15213, USA}

\author{Kenji Watanabe}
\author{Takashi Taniguchi}
\affiliation[National Institute for Materials Science]
{National Institute for Materials Science, Tsukuba, Ibaraki 305-0044, Japan}

\author{Benjamin Matthew Hunt}
\affiliation[Carnegie Mellon University]
{Department of Physics, Carnegie Mellon University, Pittsburgh, PA 15213, USA}

\author{Igor V Bondarev}
\affiliation[North Carolina Central University]
{Department of Mathematics and Physics, North Carolina Central University, Durham, NC 27707, USA}
\email{ibondarev@nccu.edu}

\author{David Snoke}
\affiliation[University of Pittsburgh]
{Department of Physics and Astronomy, University of Pittsburgh, Pittsburgh, PA 15260, USA}
\email{snoke@pitt.edu}
\phone{+86-18017854698}
\title{Charged bosons made of fermions in bilayer structures with strong metallic screening}

\abbreviations{TMDs,BEC}

\begin{document}


\begin{abstract}
  Two-dimensional monolayer structures of transition metal dichalogenides (TMDs) have been shown to allow many higher-order excitonic bound states, including trions (charged excitons), biexcitons (excitonic molecules), and charged biexcitons. We report here experimental evidence and the theoretical basis for a new bound excitonic complex, consisting two free carriers bound to an exciton in a bilayer structure. Our experimental measurements on structures made using two different materials show a new spectral line at the predicted energy with two different TMD materials (MoSe$_2$ and WSe$_2$) with both $n$- and $p$-doping, if and only if all the required theoretical conditions for this complex are fulfilled, in particular, only in the presence of a parallel metal layer that significantly screens the repulsive interaction between the like-charge carriers. Because these four-carrier bound states are charged bosons, they could eventually be the basis for a new path to superconductivity without Cooper pairing.
\end{abstract}

\vspace{1cm}

The existence of high-order complexes of charge carriers in semiconductors is well established, including biexcitons (two excitons bound together like a hydrogen molecule) and trions (an exciton plus one extra carrier) \cite{GaAstrion}. In transition metal dichalcogenide (TMD) monolayer and bilayer structures, several such states have been seen (for recent work, see, e.g., Ref.~~\onlinecite{TMDtrion}); the large exciton binding energy in TMD monolayers means that all such excitonic complexes will have a similarly large binding energy.
This raises the possibility of realizing a proposal made by V.I. Yudson two decades ago \cite{yudson} for four-carrier complexes that have a net charge, namely excitons bound to two additional charges.  These can be variously called ``tetramers''  \cite{tetra}, ``doubly charged excitons'' \cite{qdot}, or ``quaternions'' \cite{quat}; here we will use the last term. These complexes can carry current, since they have a net charge, but they are also bosonic, because they have an even number of fermions. 

A Bose condensate of such complexes would be a superfluid, and therefore also a Schafroth superconductor.\cite{schafroth}$^,$\cite{edelman}  Schafroth superconductivity based on charged excitonic complexes is a different mechanism from earlier proposals for exciton-mediated superconductivity --- in one proposal~\cite{lozo-sc}, it was argued that the presence of a magnetic field would cause neutral excitons to respond to an electric field; in another proposal~\cite{kav-sc}, exciton-polaritons were proposed to play the same role as phonons in Cooper pairing.

\begin{figure}[!h]
	\begin{center}
	\includegraphics[scale=0.5]{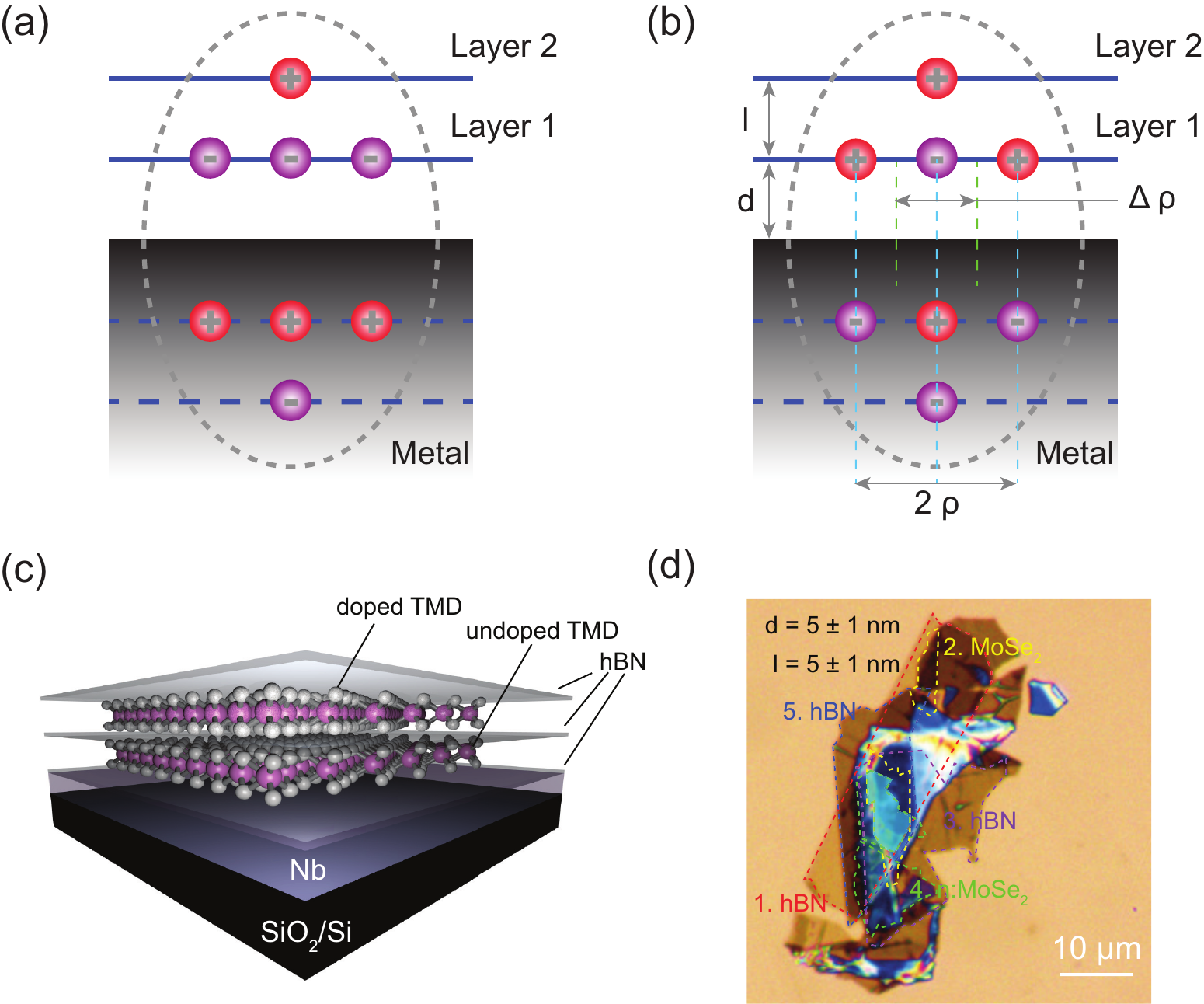}
	\caption{(a)~Quaternion geometry proposed in Ref. 3. The gray region indicates the metallic layer with image charge. (b)~Symmetric quaternion geometry considered here. See text for notations. (c)~Illustration of the fabricated structure. (d)~Image of the structure, with the layers labeled in order of deposition.}
	\label{qfig1}
	\end{center}
\end{figure}

The geometry considered by Yudson \cite{yudson} is shown in Fig.~\ref{qfig1}(a). Two semiconductor layers are placed side by side to make a bilayer structure, and this bilayer structure is placed parallel to a nearby metal layer. Under optical pumping, an exciton can be created which then picks up two free carriers. At first glance, one would not expect that a complex with three times more negative charge than positive (or vice versa) would be stable. The presence of the metal layer, however, is crucial, because it produces image charge below the surface, so that much of the repulsive interaction in the quaternion complex is canceled out.

We consider a variant of the Yudson geometry, which is structurally a trion in one layer bound to a free carrier in a parallel layer, as shown in Fig.~\ref{qfig1}(b). Our calculations and experiments, discussed below, indicate that this complex is more stable than the Yudson geometry. For the experiments, we fabricated the structure shown in Fig.~\ref{qfig1}(c), using two monolayers of transition-metal dichalgogenide (TMD) separated by insulating layers of hexagonal boron nitride (hBN).  For the TMD monolayers used here, the exciton binding energy is of order 0.2~eV~\cite{WSebinding3,WSebinding,WSebinding2,Crooker19}, depending on the dielectric constant of the surrounding material. Hexagonal boron nitride (hBN) can be used as a good insulating barrier to prevent tunneling current while still allowing Coulomb interaction between free carriers in the layers~\cite{hBN}.

 {\bf Experimental method and results}. We have created this structure for several samples, using either a bilayer made of one undoped MoSe$_2$ layer and one $n$-doped MoSe$_2$ layer, or a bilayer made of one undoped WSe$_2$ layer and one $p$-doped WSe$_2$ layer, with hBN between the monolayers in each case. This allowed us to study both cases: an exciton bound to two free electrons, and an exciton bound to two free holes. Figure \ref{qfig1}(d) shows a typical image of the stack of layers for a MoSe$_2$ structure. We have also created a large number of control samples, namely the same structures without the parallel metal layer, with different thickness of the hBN layers, with and without doping, and with just single monolayers. The thickness of the hBN layers in all cases was known to an uncertainty of $\pm 1$ nm.

\begin{figure}[!t]
	\includegraphics[scale=0.35]{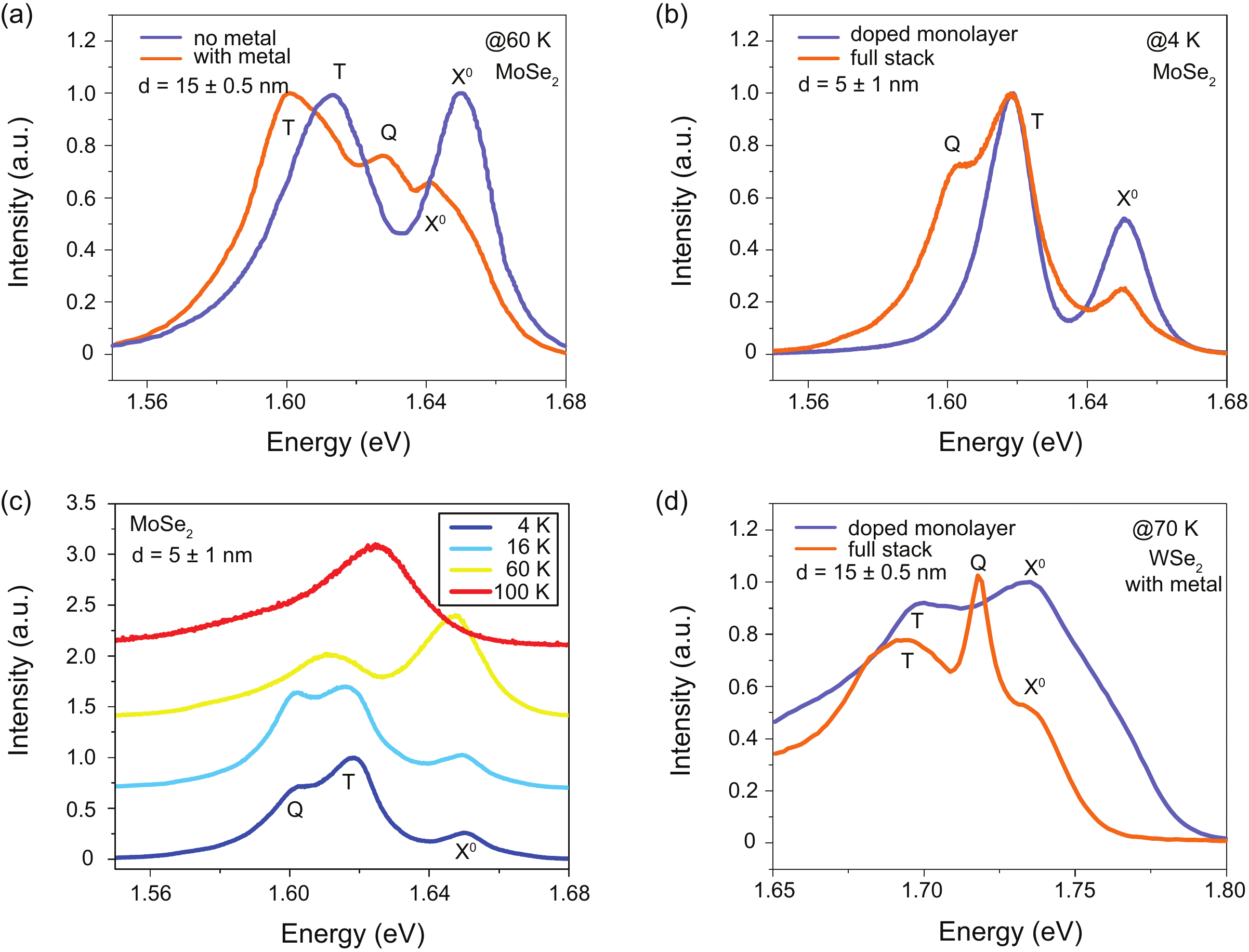}
	\caption{a) Normalized photoluminescence spectrum for the structure shown in Figure 1(c), with monolayers of MoSe$_2$ and $n$:MoSe$_2$, and $l \simeq 7$ nm and $d \simeq 15$ nm, with and without the metal layer. X$^0$ = exciton line, T = trion line, Q = proposed quaternion line.  b) The low-temperature PL spectrum from the a full bilayer structure with the metal layer, but with thinner hBN layers ($d \simeq$ 5 nm, $l \simeq d$) next to the metal, compared to one doped monolayer, with the same thickness $d$ of hBN between the monolayer and the metal.  As per the theory of Figure 3(b), the quaternion and trion lines are split just slightly when this smaller value of $d$ is used. c) Temperature dependence of the PL from the full stack sample used in (b). d) Data for a full bilayer structure with the same design as the sample used for (a), but made with monolayers of WSe$_2$ and $p$:WSe$_2$. This is compared to a monolayer of the $p$:WSe$_2$, also encapsulated in hBN, and on top of a metal layer.}
	\label{qfig2}
\end{figure}

Figure~\ref{qfig2}(a) shows photoluminescence (PL) spectra for two nearly-identical structures with and without the metal layer. We used niobium as the metal, with an hBN layer of approximately $7$ nm between the TMD layers, a spacer hBN layer of approximately $15$~nm between the metal and the first TMD monolayer, and a capping hBN layer. As seen in Figure~\ref{qfig2}, a new PL line appears, which we label Q, between the direct exciton line and the trion line. The exciton and trion energies are well known in both MoSe$_2$ and WSe$_2$, with slight variations of the order of 5-10 meV, as expected from residual strain effects \cite{sunstrain}.  As shown below, the energy of the Q line is consistent with our theoretical calculations of the quaternion binding energy. Although the absolute energy position of the exciton line fluctuates by a few meV from sample to sample and from region to region within a single sample, the relative energy differences between the exciton, trion, and quaternion lines, which are determined by the binding energies of the trion and quaternion states, are consistent with our theoretical predictions. We note that the exciton is charge neutral, and therefore, to first order, should not be affected by the presence of the metal. The trion has net charge, and so is shifted relative to the exciton line in the presence of the metal, as we observe. 

We have reproduced this effect in several samples, and have examined a number of control structures; the supplementary information file for this publication gives examples of data from several of these structures. The results can be summarized as follows:
\begin{itemize}
	\item We have reproduced the new Q line in three different WSe$_2$ structures with the same design and in three structures of similar design but with different layer thickness $d$ using MoSe$_2$.
	\item The new line appears {\em only} when the full structure is in place; it never appears when there is no metal layer, or no doping, or only a monolayer.
	\item We have performed many control experiments with monolayers that confirm the identification of all the other lines we see in WSe$_2$ and MoSe$_2$, namely the single exciton, the trion, the spatially indirect exciton, higher-order complexes such as biexcitons and charged biexcitons, and impurity lines. 
	The data from these experiments are summarized in the Supplementary file.
	\item We have performed pump-power dependent studies in both WSe$_2$ and MoSe$_2$ that show that the new line increases linearly with pump power (see, e.g., Figures S7 and S8 of the supplementary file) just as the exciton does, which rules out that it is a biexciton line or an impurity line. Biexciton lines have superlinear behavior with density, as they need two photons to be created; impurity lines have sublinear behavior with increasing pump power.
	\item With both the MoSe$_2$ and WSe$_2$ structures, we have performed temperature-dependent measurements like those shown in Figure~\ref{qfig2}(c) which show that the Q line is visible from 4 K up to around 60-70 K in both types of structure; the trion has similar behavior. The temperature dependence of the trion and quaternion lines can be understood by several effects. First, for a quaternion to be formed, an exciton must find two free electrons (or holes), which means that their relative numbers will be determined by a mass-action equation \cite{sscion}. Second, the number of free carriers will change as a function of temperature; at low temperature, these carriers will mostly be bound to impurities, and therefore the trion and quaternion intensities will drop. Third, at high temperature, all of the PL lines undergo thermal broadening, which makes it hard to distinguish one line from another.
\end{itemize}

 {\bf Theoretical model}. These results, and the identification of the Q line as a quaternion, are consistent with the straightforward theory based on the configuration space method of the binding energy calculation~\cite{Bondarev2016,Bondarev11PRB,Bondarev14PRB,BondVlad18}. The same method gives experimentally confirmed binding energies both for interlayer trions~\cite{Science} and for biexcitons~\cite{JonFinley} in TMD materials. The configuration-space theory has also been able to explain the evidence for a positive/negative trion binding energy difference~\cite{Science,preprint-bond}. We use it here with the additional inclusion of the image charges in the metal layer. The approach itself was originally pioneered by Landau~\cite{LandauQM}, Gor'kov and Pitaevski~\cite{Pitaevski63}, Holstein and Herring~\cite{Herring,Herring2} in their studies of molecular binding and magnetism.

In our model, shown in Fig.~\ref{qfig1}(b), the intralayer (direct) trion makes the ``core'' to attach a like charge from the other monolayer to form the quaternion complex. The axial symmetry of such a complex relative to the axis perpendicular to the bilayer, supplemented by the image charges of the same symmetry in metal, makes its ground-state coordinate wave function even (no nodes), as required for stability.\cite{LandauQM}  

The first step is to calculate the binding energy of this intralayer trion in the presence of the image charges. This calculation follows the same method as used elsewhere for charged indirect excitons, discussed in the Supplementary file, which treats the trion as a bound state of two equivalent excitons sharing the same hole (or electron)~\cite{Bondarev2016}. The trion bound state can then be found as the minimum in the energy landscape for the two configurations of the free carrier belonging to either exciton, with a tunneling barrier between the two configurations. The tunneling exchange rate $J_{X^\pm}$ controls the binding strength. The binding energy of the trion ground state is given by~\cite{preprint-bond}
\begin{equation}
	E_{\!X^{^{\pm}}}(\sigma,r_0)=-J_{\!X^{^{\pm}}}(\Delta\rho=\!\Delta\rho_{\!X^{^{\pm}}},\sigma,r_0),
	\label{Eb}
\end{equation}
where the electron-hole mass ratio $\sigma\!=\!m_e/m_h$ and the electrostatic screening length $r_0$ are the intrinsic parameters of the monolayer, and $\Delta\rho_{X^\pm}$ is the equilibrium exciton center-of-mass-to-center-of-mass distance in the trion (obtained variationally to maximize the tunneling exchange). We use the Keldysh-Rytova interaction potential for the charges confined in the monolayer to properly account for the screening effect~\cite{KeldyshRytova,Rytova}; its screening length $r_0\!=\!2\pi\chi_{2D}$ where $\chi_{2D}$ is the in-plane polarizability of the 2D material~\cite{Rubio11,Berkelbach2013}. The 3D ``atomic units''\space are used~\cite{preprint-bond,LandauQM,Pitaevski63,Herring,Herring2} with distance and energy expressed in units of the exciton Bohr radius $a^\ast_B\!=\!0.529\,\mbox{\AA}\,\varepsilon/\mu$ and Rydberg energy $Ry^\ast\!\!=\!\hbar^2/(2\mu\,m_0a_B^{\ast2})\!=\!13.6\,\mbox{eV}\,\mu/\varepsilon^2$, respectively, where $\varepsilon$ is the effective average dielectric constant of the structure and $\mu\!=\!m_e/(\lambda\,m_0)$ with $\lambda\!=\!1+\sigma$ is the exciton reduced effective mass (in units of the free electron mass $m_0$). The explicit form of $J_{\!X^{^{\pm}}}$ and more theory details can be found in the SI file.


In the presence of a metal, the total potential energy of the intralayer trion is
\begin{eqnarray}
	U(\rho,d) =U_0(\rho)+2\left( \frac{4}{\sqrt{(2d)^2+\rho^2}}-\frac{2}{\sqrt{(2d)^2+(2\rho)^2}} - \frac{3}{2d}\right),
	\label{Ut}
\end{eqnarray}
where $\rho$ is the in-plane distance between the hole and the electron shown in Fig.~\ref{qfig1}(b), and $U_0(\rho)$ is the electron-hole potential interaction energy in the absence of a metal already included in Eq.~(\ref{Eb}). The second term comes from the image charge interaction with $d$ being the distance of the monolayer from the metal (the distance between the image and the original). For the quaternion, in a similar manner, the total potential energy with the image charge interaction included is
\begin{eqnarray}
	U(\rho,d,l) &=& U_0(\rho)+2\left(\frac{4}{\sqrt{(2d)^2+\rho^2}} -\frac{2}{\sqrt{(2d)^2+(2\rho)^2}}-\frac{3}{2d}\right.\nonumber\\
	&& \left. +\frac{2}{\sqrt{l^2+\rho^2}}-\frac{1}{l}+\frac{2}{2d+l} -\frac{1}{2d+2l} - \frac{4}{\sqrt{(2d+l)^2+\rho^2}}\right),
	\label{Uq}
\end{eqnarray}
where $l$ is the thickness of the spacer layer between the two TMD monolayers --- see Fig.~\ref{qfig1}(b).

The PL emission spectra in Figure~\ref{qfig2} can be understood in terms of Eqs.~(\ref{Eb})--(\ref{Uq}). The PL photon energy is given by the initial energy minus the final energy. In the exciton recombination process, the final state is nothing, so the energy of the photon emitted is the bandgap minus the exciton binding energy in the presence of a metal. For the intralayer trion, the final state is a single electron (or hole), which in the presence of a metal has the energy $2\,[-1/(2d)]$ due to the image-charge interaction. For the quaternion, there are two final electrons (or two holes), and so the final energy is $2\,[-1/(2d)\!-\!1/(2d\!+\!2l)]$. Subtracting these final state energies, together with $U_0(\rho)$, from $U(\rho,d)$ and $U(\rho,d,l)$ in Eqs.~(\ref{Ut}) and (\ref{Uq}), respectively, and adding the intralayer trion binding energy with no metal present of Eq.~(\ref{Eb}), we obtain the recombination energies of interest in the presence of a metal as functions of $d$ and $l$.


\begin{figure}[t]
	\includegraphics[scale=0.82]{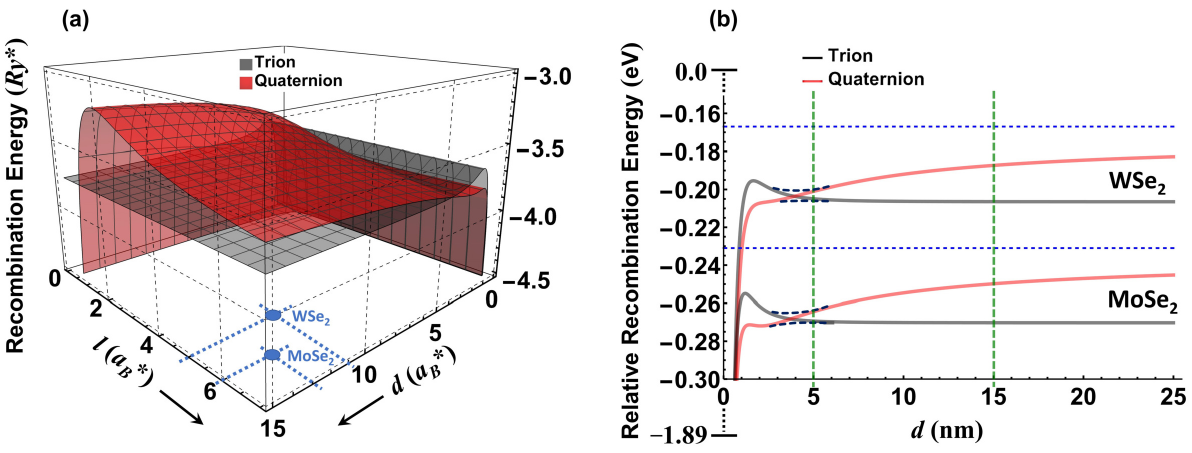}
	\caption{(a) The recombination energies for the intralayer trion and quaternion as functions of $d$ and $l$ in atomic units as given by Eqs.~(\ref{Eb})--(\ref{Uq}) for $\sigma\!=\!1$ and $r_0\!=\!0.05$. The blue spots show our experimental $d$ and $l$ in these units, determined with $\varepsilon\!=\!5.87$ for WSe$_2$ and MoSe$_2$ as described in the text. (b)~The trion and quaternion recombination energies for WSe$_2$ and MoSe$_2$ relative to their respective bandgaps. The energies are calculated as functions of $d$ in nanometers for the experimental value of $l\!=\!7$~nm, with $\varepsilon\!=\!5.87$ and $e$-$h$ reduced masses $0.23$~(WSe$_2$) and $0.27$~(MoSe$_2$)~\cite{WSe2mass}. The horizontal dotted blue lines trace the exciton binding energies of $167$~meV~(WSe$_2$) and $231$~meV~(MoSe$_2$)~\cite{Crooker19}; the trion binding energy of $40$~meV for both WSe$_2$ and MoSe$_2$ is used. The vertical dashed green lines trace the approximate values of $d$ used in the experiments shown in Fig.~\ref{qfig2}(a) and (b). As discussed in the text, an anticrossing of the Q and T lines is expected, illustrated by the black dashed curves.}
	\label{qfig3}
\end{figure}

Figure~\ref{qfig3}(a) shows the recombination energies for the intralayer trion and quaternion as functions of $d$ and $l$ calculated in atomic units with the screening parameter $r_0=0.05$ as an example. As seen in this plot, when $d$ is very small, the quaternion recombination energy shifts to below that of the trion. Near-zero values of $d$ correspond to directly placing the bottom TMD layer on the metal surface. This would suppress the formation of the trion complex, and hence quaternion formation, due to strong screening. Such behavior is seen in the sudden drop at $d \sim 0$ for both energies. 
The quaternion recombination energy slowly increases as $d$ increases and down as $l$ increases, to eventually exceed the exciton recombination energy for $d$ large enough ($d\!=\!\infty$ is the no-metal case) and to approach the trion recombination energy for $l$ large enough, which would make it unstable to conversion down to excitons and trions, respectively. Thus, the intervening hBN layer thickness plays a crucial role. We see all three recombination lines individually because our experimental conditions are different from those of Ref.~~\onlinecite{WSebinding2}, for example, where a TMD monolayer was placed directly onto a metal.

Figure~\ref{qfig3}(b) shows the relative recombination energies for the trion and quaternion in eV, as functions of the distance $d$ in nanometers. This figure is the cross-section of Fig.~\ref{qfig3}(a) converted to physical units with $l\!=\!7$~nm and shifted vertically by choosing the appropriate $r_0$ to match the $1.89$~eV WSe$_2$ bandgap with the exciton binding energy of $167$~meV and the $1.874$~eV MoSe$_2$ bandgap with the exciton binding energy of $231$~meV reported recently from precision experiments with TMD monolayers embedded in hBN \emph{without} metal~\cite{Crooker19}, which the theory here nicely reproduces, and also to match the $40$~meV intralayer trion binding energy we observe both for WSe$_2$ and for MoSe$_2$ samples. The figure is obtained for $\varepsilon\!=\!5.87$ assuming that the static dielectric permittivity of our entire system is dominated by the dielectric permittivity of the bulk hBN-material, that is $\varepsilon\!=\![2\,\varepsilon_{hBN}(\mbox{in-plane})\!+\!\varepsilon_{hBN}(\mbox{out-of-plane})]/3$, where $\varepsilon_{hBN}(\mbox{in-plane})\!=\!6.93$ and $\varepsilon_{hBN}(\mbox{out-of-plane})\!=\!3.76$ as was reported recently from advanced numerical simulations~\cite{hBNdiel}. We used the exciton reduced masses $\mu\!=\!0.23$ and $0.27$ for WSe$_{2}$ and MoSe$_{2}$, respectively, neglecting slight $e$-$h$ effective mass differences~\cite{WSe2mass}, to give us $\sigma\!=\!1$ with equal binding energies for the positive and negative trions (see the SI file for more details). With these material parameters we obtain $a_B^{\ast}\!=\!1.35$~nm, $Ry^{\ast}\!=\!0.09$~eV and $a_B^{\ast}\!=\!1.15$~nm, $Ry^{\ast}\!=\!0.11$~eV for our WSe$_2$ and MoSe$_2$ samples, respectively. Then $l\!=\!7$~nm corresponds to $5.18$ in atomics units for the WSe$_2$ samples and $6.09$ in atomic units for the MoSe$_2$ samples. We use these $l$ values in Eq.~(\ref{Uq}), whereby the set of Eqs.~(\ref{Eb})--(\ref{Uq}) with $r_0$ fixed provides the crosscuts of Fig.~\ref{qfig3}(a) as functions of $d$ in atomic units. By using $a_B^{\ast}$ and $Ry^{\ast}$ obtained, these functions can be converted to physical units, followed by adjusting $r_0$ to vertically shift the curves to match the respective WSe$_2$ and MoSe$_2$ bandgap energy patterns. We find $r_0\!=\!0.0162$ and $0.0303$ in atomic units for the WSe$_2$ and MoSe$_2$ samples, respectively. We note that our $r_0$ parameter represents the screening of the like-charge carriers forming the intralayer trion, which are therefore separated by distances at least of the order of $2a_B^{\ast}$, whereby our $r_0$ is contributed more by surrounding hBN material than by TMD itself and so it can be quite different from order-of-magnitude greater values reported experimentally for excitons in hBN-embedded TMD monolayers~\cite{Crooker19}. The low $r_0$ values we obtain are consistent with the experimental reports of the exciton emission linewidth approaching the homogeneous limit for TMD monolayers embedded in between thick hBN layers~\cite{XMarie17}. We checked that increasing $\varepsilon$ by a factor as large as $2.5$ makes no significant change to the physical picture presented in Fig.~\ref{qfig3}, showing its robustness.

From Figure~\ref{qfig3}(b) one can see that increasing $d$ pushes the Q line up to cross over the X$^0$ line, thus making the quaternion state unstable. Decreasing $d$, on the other hand, shifts the Q line down where it can be seen to intersect with and eventually shift below the T line. We emphasize, however, that according to a general theorem of quantum mechanics~\cite{LandauQM}, for Hamiltonians with an external parameter (whereby their eigenvalues are functions of that parameter as well), only states of different symmetry can intersect, while the intersection of like-symmetry states is prohibited. In our case here, the trion and quaternion states have the same axial and reflection symmetry as per Fig.~\ref{qfig1}(b). We therefore expect that the Q and T line crossing we obtain within the classical image-charge approach will turn into an anticrossing of the Q and T lines as the parameter $d$ is varied, when quantum mechanics is taken into account. The details of this must be treated elsewhere; here we show this anticrossing behavior by the black dashed curves in Fig.~\ref{qfig3}(b), with vertical green dashed lines tracing our respective experimental $d$ values as per the PL data presented in Figs.~\ref{qfig2}(a) and (b). The Q line is seen to shift towards the higher energy for thicker $d$ values (15 nm versus 5 nm in our experiments) just as our theory predicts. We note that the shift of the Q line down in energy as $d$ is decreased, leading to the tiny splitting $\sim\!10$~meV of the Q and T lines at $d=5$~nm, was a prediction of our theory before the experiment with $d=5$~nm was done, and not a retrodiction.

 {\bf Conclusions and outlook}. Our experimental observations are strong evidence for these doubly-charged excitonic complexes, or quaternions, in bilayer TMD structures near metallic layers, fully consistent with the quantitative theoretical predictions for their existence. Our theory shows this complex to be robust to material parameter variation, and the trend of the shift of the quaternion line with distance from the metallic layer is fully consistent with the theory. All of the other lines which we observe spectroscopically can be identified as known lines, as discussed in the Supplementary Information. Full confirmation and the identification of the quaternion states will depend on the observation of the direction of their motion in response to an in-plane electric field, since they have net charge. This can be accomplished, for example, by optical measurements of drift, akin to that used to confirm the existence of trions \cite{GaAstrion}. Using a top gate to create a vertical electric field to vary the free carrier density is not straighforward in these structures. The stability of the quaternions depends crucially on the layer asymmetry of the entire plane-parallel semiconductor-metal structure. Another conducting layer (the top gate) would make the structure nearly symmetric by adding extra image charges on the top, which would greatly diminish the binding energy of the quaternion states. 

Like a Bose condensate of excitons, a Bose condensate of quaternions would be metastable to recombination and require optical pumping for steady state. But as the burgeoning field of experimental and theoretical work on Bose condensates of exciton-polaritons has shown~\cite{devead,balili,deng,caru,PT}, such a steady-state optical pumped system can indeed undergo condensation, including the effects of superfluidity, and can reach equilibrium in the steady-state with a well-defined temperature~\cite{PRL-therm,marz-equil}. The quaternion particles discussed here do not have a polariton nature, and therefore are more similar to pure exciton condensates, such as interlayer excitons in bilayer systems  \cite{comb-review,lozo-review,snoke-review}, which are subject to much greater disorder effects. However, since the quaternions have charge, they will have much stronger interactions, which may cause a condensate of such particles to be more readily in the Thomas-Fermi regime, with a common chemical potential which smooths out the disorder effects.

Experiments with high excitation intensity could push the quaternion density high enough for condensation, but at high density, nonradiative collisional Auger recombination may become important. 
Although true BEC is not possible in one and two dimensions for noninteracting bosons~\cite{Hohenberg67}, it is known that a slight lateral confinement enables BEC for noninteracting bosons both in 2D and in 1D,\cite{Bagnato91Dai02,BondMel14PRB} and in general, in any finite system in which the size of the system is small compared to the coherence length, the system can undergo a transition indistinguishable from BEC.  Our results here indicate that quaternion physics in bilayer systems with metal layers is a promising field of research.

\begin{acknowledgement}

This work was supported by the US Army Research Office under MURI award W911NF-17-1-0312 (Z.S., J.B, Q.W., H.A., N.H., J.C., and D.W.S.) and by the U.S. Department of Energy, Office of Science, Office of Basic Energy Sciences, under award number DE-SC0007117 (I.V.B). Z.S. also acknowledge support from Joint Physics Research Institute Challenge Grant of the NYU-ECNU Institute of Physics at NYU Shanghai. K.W. and T.T. acknowledge support from the EMEXT Element Strategy Initiative to Form Core Research Center, Grant Number JPMXP0112101001 and the CREST(JPMJCR15F3), JST. B.M.H. and Q.C. were supported by the Department of Energy under the Early Career Award program (\#DE-SC0018115).

\end{acknowledgement}

\begin{suppinfo}

\begin{itemize}
  \item Supplementary Information includes 1. Temperature-dependent spectra of all the other control samples; 2. Power-dependent experiment for both TMDs based samples; 3. Theory for Trion Binding Energy Calculations.
\end{itemize}

\end{suppinfo}


\end{document}


\newpage
	
\section{I. Experimental Photoluminescence Spectra}

\begin{figure}[h]
	\begin{center}
		\includegraphics[width = .65\linewidth]{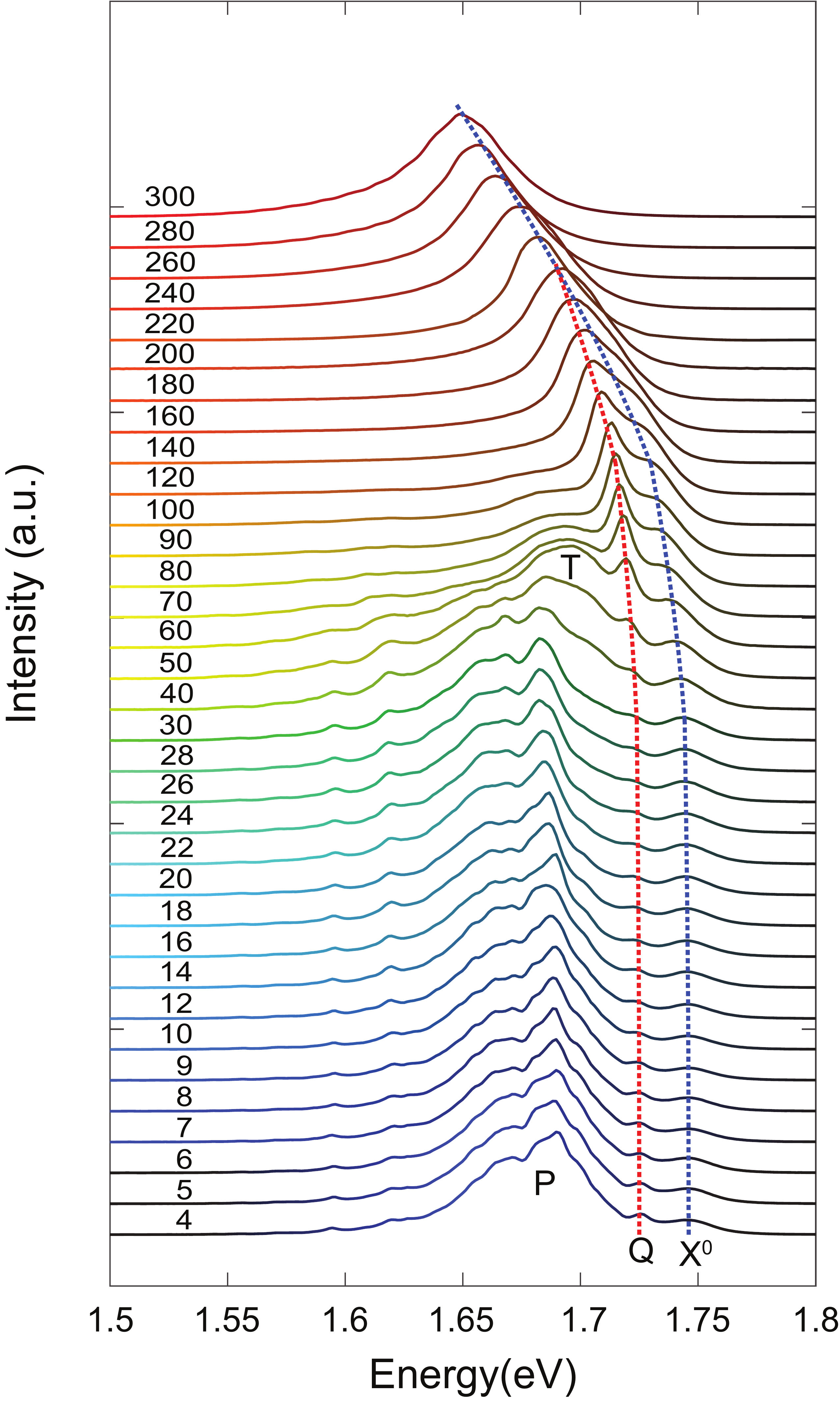}
	\end{center}
	\caption{Normalized photoluminescence spectrum for a full structure according to the design of Figure 1(c) of the main text, but with undoped WSe$_2$ and $p$-doped WSe$_2$ monolayers, with $d \simeq 15$ nm and $l \simeq 8$ nm, at various temperatures. The dashed lines are fits to the Varshni equation for band gap shift of the lines. X$^0$ = exciton, T = trion, P = impurity lines, and Q = the candidate for the quaternion emission. The doping density for the p:WSe$_2$ is $\sim$10$^{17-18}$ cm$^{-3}$ (Nb dopant).
	}
	\label{sifig0}
\end{figure}

\begin{figure}[h]
	\begin{center}
		\includegraphics[width = .65\linewidth]{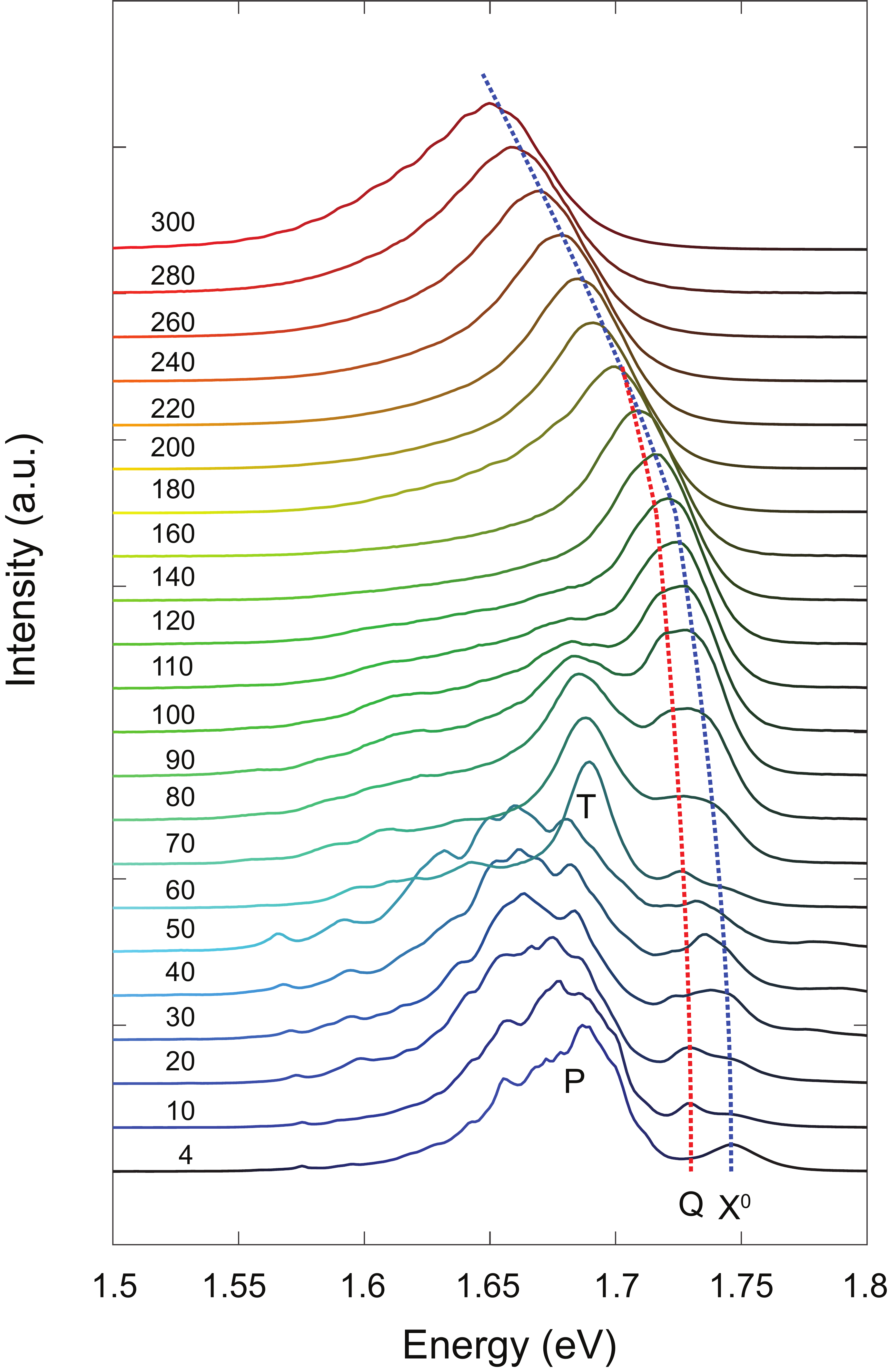}
	\end{center}
	\caption{Second sample with the same structure as the sample used for Figure S1. The direct exciton is seen at about the same energy, equal to 1.747 eV at low temperature, and the quaternion line is also clearly identifiable at low temperature. The trion line (seen most clearly at around $T = 100$ K) also appears at the same energy, about 40 meV below the exciton line.  The dashed lines are fits to the shift of the lines predicted by the Varshni formula for band gap shift. }
	\label{sifig1}
\end{figure}

\begin{figure}[h]
	\begin{center}
		\includegraphics[width = .65\linewidth]{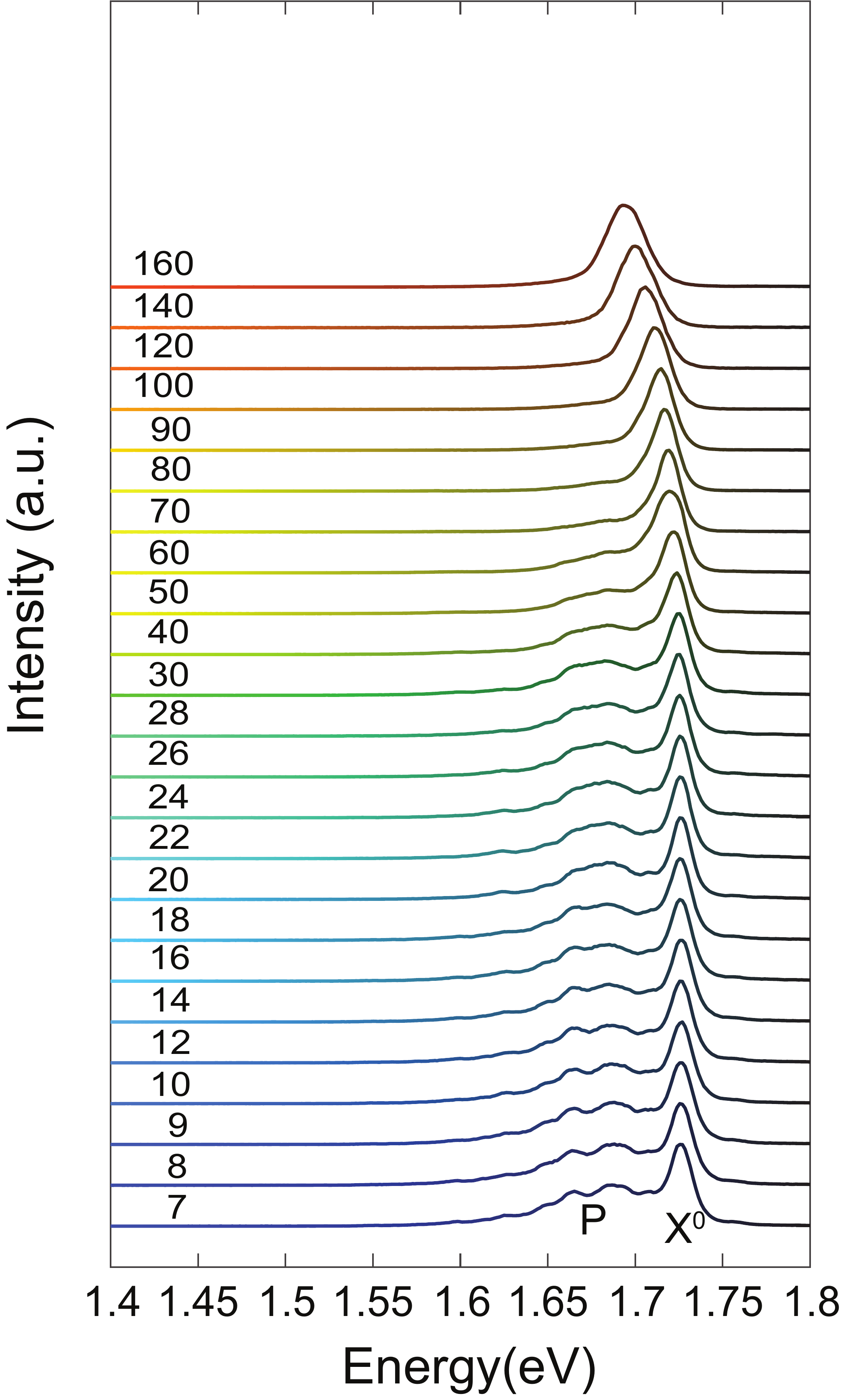}
	\end{center}
	\caption{Control sample \#1, consisting of one undoped monolayer of WSe$_2$ encapsulated in hBN, with no metal layer. The direct exciton line dominates the whole spectrum. Its energy is about 15 meV lower than for all samples with a metal layer, indicating slightly stronger exciton binding when there is no image charge of the metal. No quaternion line is seen, and trion emission is either absent or buried in the impurity emission.}
	\label{sifig2}
\end{figure}

\begin{figure}[h]
	\begin{center}
		\includegraphics[width = .65\linewidth]{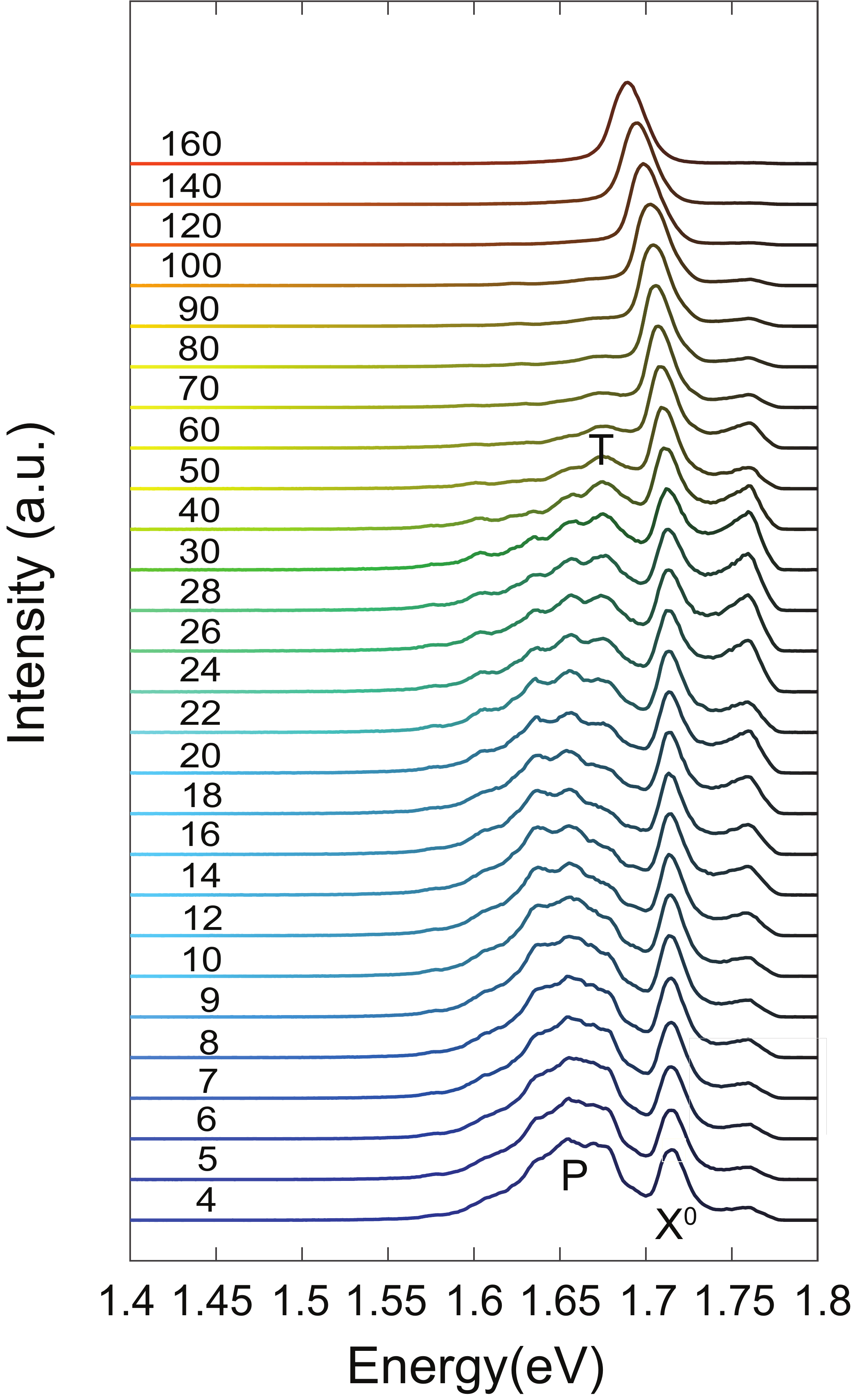}
	\end{center}
	\caption{Control sample \#2, consisting of one p-doped monolayer of WSe$_2$ encapsulated in hBN, with no metal layer. The direct exciton line appears at the same energy as control sample \#1, which also had no metal layer. The peak above 1.75 eV is an artifact of scattered light cut off by a 700-nm long-pass filter.}
	\label{sifig3}
\end{figure}

\begin{figure}[h]
	\begin{center}
		\includegraphics[width = .65\linewidth]{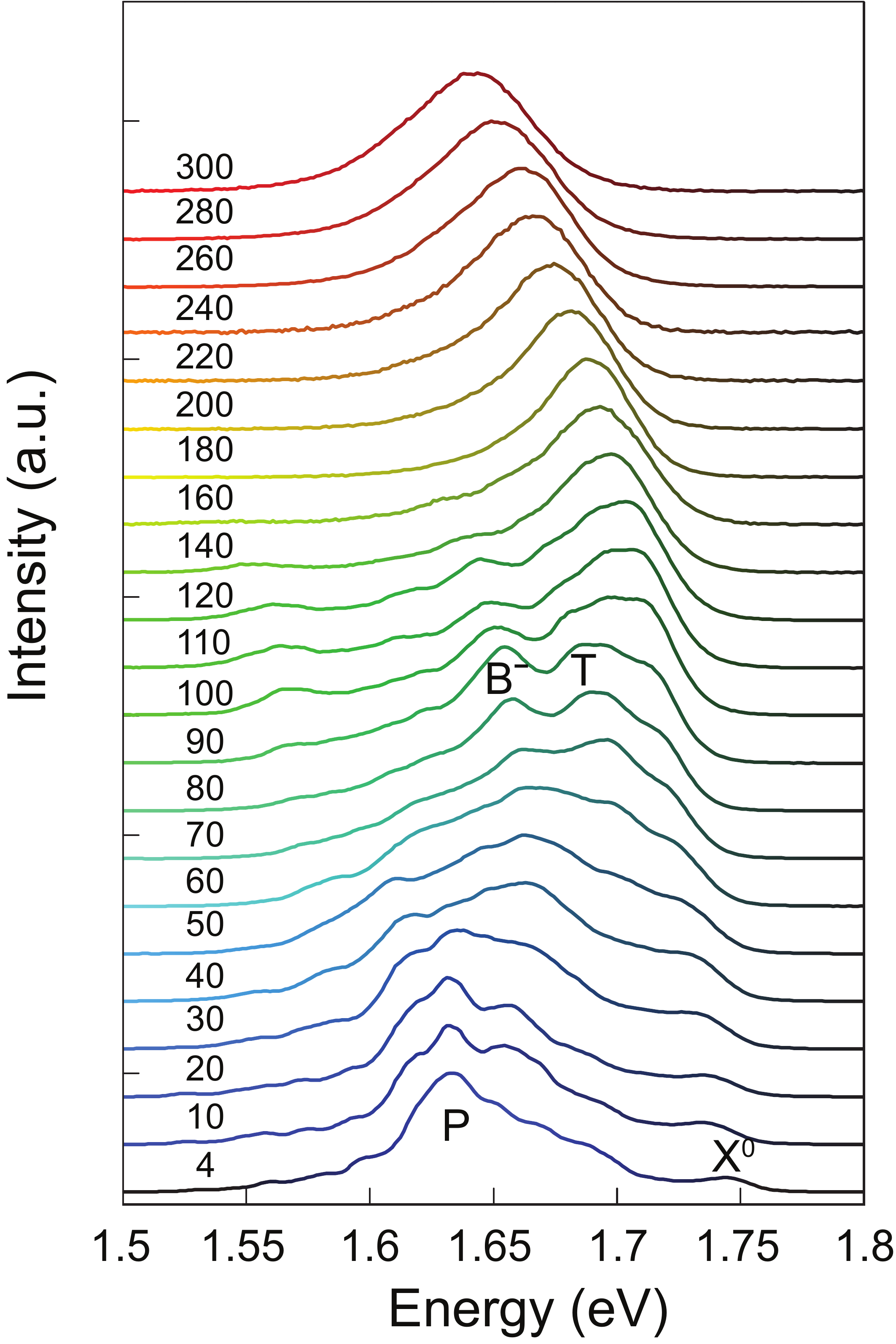}
	\end{center}
	\caption{Control sample \#3, consisting of one p-doped monolayer of WSe$_2$ in the presence of a metal layer. The thickness of the hBN layer between the monolayer and the metal was 15 nm. The direct exciton and trion lines appear at the same energies as for the bilayer stacks in the presence of a metal layer; a charged biexciton line (B$^-$) is visible at 1.65 eV at $T = 100$ K. No quaternion line is seen.}
	\label{sifig4}
\end{figure}

\begin{figure}[h]
	\begin{center}
		\includegraphics[width = .65\linewidth]{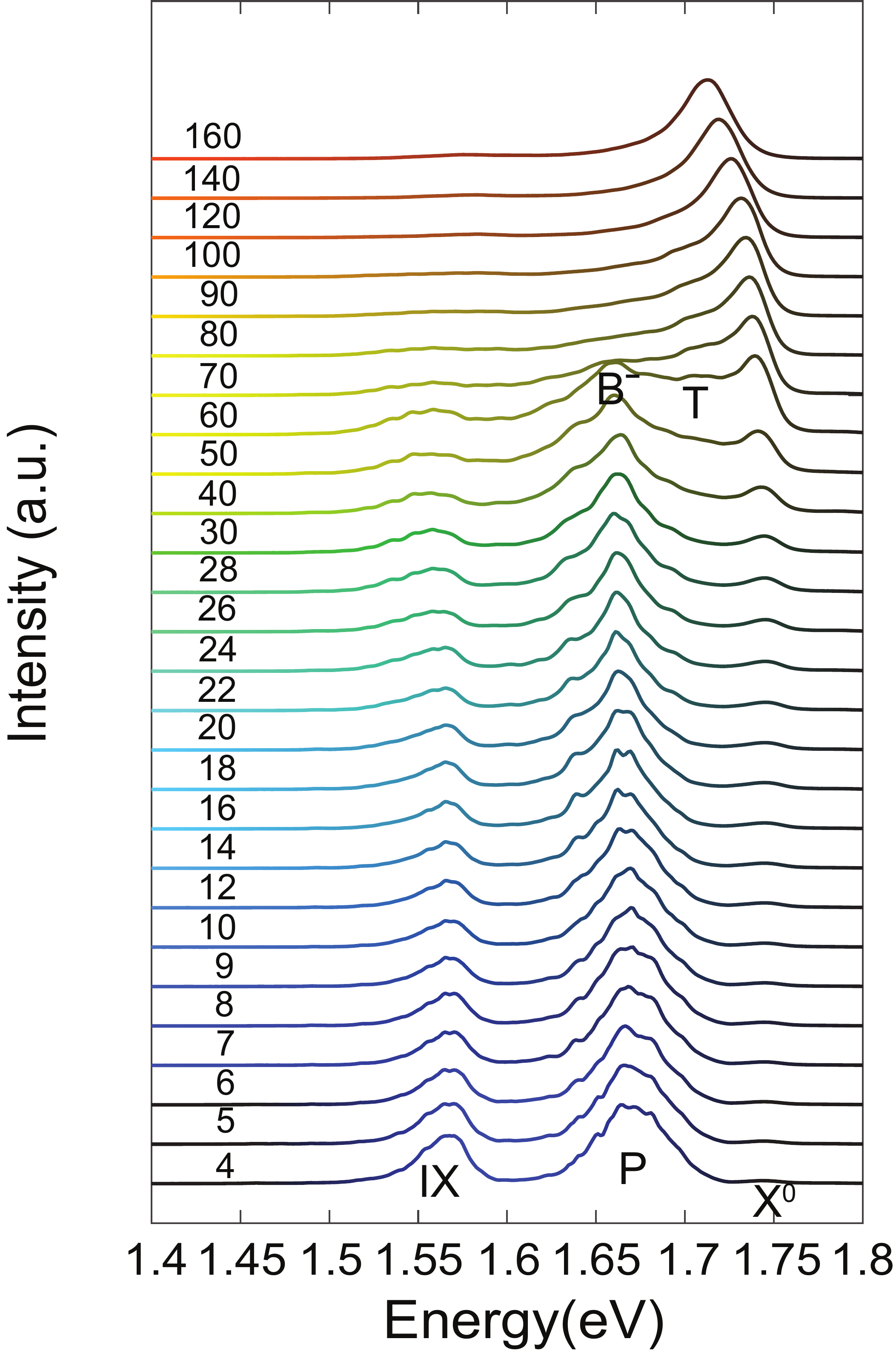}
	\end{center}
	\caption{A second WSe$_2$-based sample with the same structure as the sample discussed in the main text, but without the metal layer, and with the thickness $l$ of the hBN layer between the monolayers much thinner, approximately 2 nm. The direct exciton is seen at the same energy position, close to 1.75 eV at low temperature, but the quaternion line is not seen. An interlayer exciton (IX) is seen prominently at 1.57 eV at low temperature, while the trion line is suppressed or buried in the impurity PL, and the quaternion line is not seen. This sample, and other measurements of its PL properties, such as time-resolved PL showing a long lifetime for the IX, is discussed in another publication~\cite{IXpaper}. }
	\label{sifig5}
\end{figure}

{\bf Extended study of WSe$_2$ structures}. Figures \ref{sifig0} to \ref{sifig5} show the PL spectra of various samples as temperature is varied. The thickness of all hBN layers was approximately 15 nm for all of these cases.  The curves are labeled by the bath temperature in each case. In all figures,  X$^0$ indicates the direct exciton line, P indicates impurity lines, T indicates the trion line, Q indicates the proposed quaternion line, and IX indicates the spatially indirect exciton line. In Figures \ref{sifig4} and \ref{sifig5}, B$^-$ indicates a charged intralayer biexciton line, based on the identification of Ref.~~\onlinecite{biexnew2}.

The data of Figures \ref{sifig0} and \ref{sifig1} are fit to the Varshni formula for the exciton energy shift with temperature. This shift is due to the many-body renormalization of the states due to phonon interaction (see Section 8.3 of Ref.~~\onlinecite{snokebook}):
\[
E(T) = E(0) - \frac{\alpha T^2}{T+\beta}\,,
\]
where $\alpha$ is a parameter giving the interaction strength of the state under consideration with optical phonons, and $\beta$ is a parameter approximately equal to the average optical phonon energy in units of Kelvin. For the fits shown, we used $\beta = 328$~K, which is consistent with the measured optical phonon energy in WSe$_2$ of 31~meV~\cite{WSe2phonon}, and $\alpha = 0.00069$ for the exciton state and $\alpha = 0.00043$ for the quaternion state. For the data of Figure 2 of the main text, $E(0) = 1.746$~eV for the exciton and $E(0) = 1.725$~eV for the quaternion, while for the data of Figure S1 here, $E(0) = 1.7463$~eV for the exciton and $E(0) = 1.73$~eV for the quaternion, reflecting the effect of slightly different dielectric constant for the environment of the layers.

\begin{table}[ht]
	\centering
	\label{tab:table1}
	\resizebox{\textwidth}{!}{\begin{tabular}{|c|cc|c|c|c|c|c|}
			\hline
			& doped bilayer  &  &undoped & doped  & doped monolayer & doped bilayer & Ref.~\protect[1]\\
			& with metal & & monolayer  & monolayer &  with metal &  without metal & \\
			& (Sample 1) & (Sample 2) &&&& &\\
			\hline
			X\textsuperscript{0} & 1.746 & 1.747 & 1.725 & 1.720 & 1.745 & 1.745 & 1.745\\	\hline
			Q & 1.725 & 1.730 & --- & --- & --- & --- & --- \\   	\hline
			T & 1.70 & 1.70 & --- & 1.68 & 1.69 & 1.70 & 1.71\\ \hline  						
			P & 1.6-1.7 & 1.6-1.7 & 1.6-1.7 & 1.55-1.7 & 1.55-17 & 1.6-1.7 & 1.6-1.7\\ \hline
			B$^-$ & --- & --- & --- & --- & 1.651$^*$ & 1.66$^{**}$ & 1.68\\
			\hline
	\end{tabular}}
	\caption{Measured values of the various excitonic complex spectral lines for WSe$_2$, in electron-Volts (eV), at $ T = 4$ K. X$_0$ = direct intralayer exciton; T= intralayer trion; Q = quaternion candidate; P = impurity lines (the total spectral range is given); B$^-$ = charged biexciton ($^*$data at $T = 100$ K; $^{**}$data at $T = 60 $ K). Data from Ref.~\protect[1] are for $T = 10$ K.}
	\label{tablelines}
\end{table}

Table~\ref{tablelines} summarizes the peak energies of the spectral lines we observe in WSe$_2$ at low temperature. The results in WSe$_2$ can summarized as follows:
\begin{itemize}
	\item Impurity lines are seen even without the metal layer (see Figs.~\ref{sifig2}, \ref{sifig3}, and \ref{sifig5}) at energies in agreement with previous work on 2D monolayers~\cite{Wang2018}, and occur only at low temperature.
	\item The intralayer trion line showing up at $1.7$~eV at elevated temperatures is well identified in other works \cite{Arora2015} and occurs separately from the new line, and coexists with it (e.g., at $T\!\sim\!80-\!100$~K in Fig.~2 of the main text).
	\item Intralayer biexcitons, formed from two excitons both in the same layer, were initially thought to occur at much lower energy, in the same range as the impurity lines \cite{biex}, but recent work \cite{biexnew1,biexnew2,biexnew3,biexnew4} has observed evidence for an intralayer biexciton line much nearer to the direct exciton line, in the same spectral range as our quaternion line. However, in those experiments, the biexciton line was seen in monolayers, and had the telltale dependence of intensity going as the square of the exciton intensity, which is expected since two excitons must collide to form an biexciton. In our experiments, the quaternion line is never seen in a monolayer, and has a linear dependence of its intensity, as discussed below. The reason for the difference in the experiments is that the experiments showing the biexciton line took great care to have undoped layers with negligible free carrier density, while ours used a $p$-doped layer. It is actually quite difficult to observe intralayer biexcitons in WSe$_2$, and most experiments do not observe this line. In the presence of free carriers, the intralayer biexiton is unstable toward formation of charged biexcitons, discussed below. 
	
	\item Surprisingly, there is evidence \cite{biexnew2} of stable {\em five}-carrier complexes, made of two excitons plus one extra charge. This shows that the idea of a family of excitonic complexes with 2, 3, 4, and even 5 carriers in a bound state is now standard for TMD materials. The charged biexciton in WSe$_2$ shows up in monolayers, and is at much lower energy than the intralayer exciton, around 1.65 eV. We see evidence for this line in Figures \ref{sifig4} and \ref{sifig5}.
	
	\item Interlayer exciton lines occur at much lower energy due to the band offset between the layers, because the PL photon emitted comes from an interlayer recombination process, as confirmed by other works~\cite{Science} and by our own work with related samples (see Fig.~\ref{sifig5} and Ref.~~\onlinecite{IXpaper}. In bilayer structures with and without a metal layer, the appearance of an indirect (interlayer) exciton line is a strong function of the thickness of the hBN layer between the TMD layers; for a $2$-nm layer the indirect exciton line appears prominently (Fig.~\ref{sifig5}), while for a $7$-nm layer, as used in the structure of Fig.~1(d) of the main text, there is no discernible indirect exciton line. The identification of the indirect exciton line in this and other samples was confirmed by lifetime measurements showing it has much longer lifetime than the direct exciton~\cite{IXpaper}.
	\item Charged interlayer excitons, which we do not see, consisting of two holes in one monolayer and one electron in the other monolayer (or vice versa), require a thinner interlayer spacer to be visible spectroscopically ($l \sim\!1$~nm as opposed to $5-7$~nm we have in the samples used for Fig.~2 of the main text). They will also have the PL energy shifted down, below the interlayer exciton line, since their PL process requires an interlayer exciton to recombine. They also have small binding energies ($\sim\!10\!-\!15$~meV demonstrated both theoretically and experimentally~\cite{BondVlad18,Science}).
	\item Interlayer biexcitons require high excitation power and so far were only observed in laterally confined TMD bilayers~\cite{JonFinley} due to their vanishingly small binding energy~\cite{BondVlad18}. They would not be stabilized by the presence of a metal either because of their overall electrical neutrality.
	\item The trion state formed by the exciton in one monolayer and a hole (or electron) in the other monolayer cannot show up in our spectra as such a state possesses no intrinsic axial symmetry necessary for it to be stable. From general quantum mechanics, the ground state of a stable quantum system must have no nodes~\cite{LandauQM}. For a few-particle complex to be stable, its coordinate wave function has to be even (no nodes), which in our case can only occur if the complex has an axially symmetric charge distribution relative to the axis perpendicular to the bilayer. This main feature of the quaternion complex we claim to observe is totally absent from such a trion complex. Adding a metal does not change the symmetry and so does not help stabilize this trion state.
\end{itemize}

\begin{figure}[h]
	\begin{center}
		\includegraphics[width = .65\linewidth]{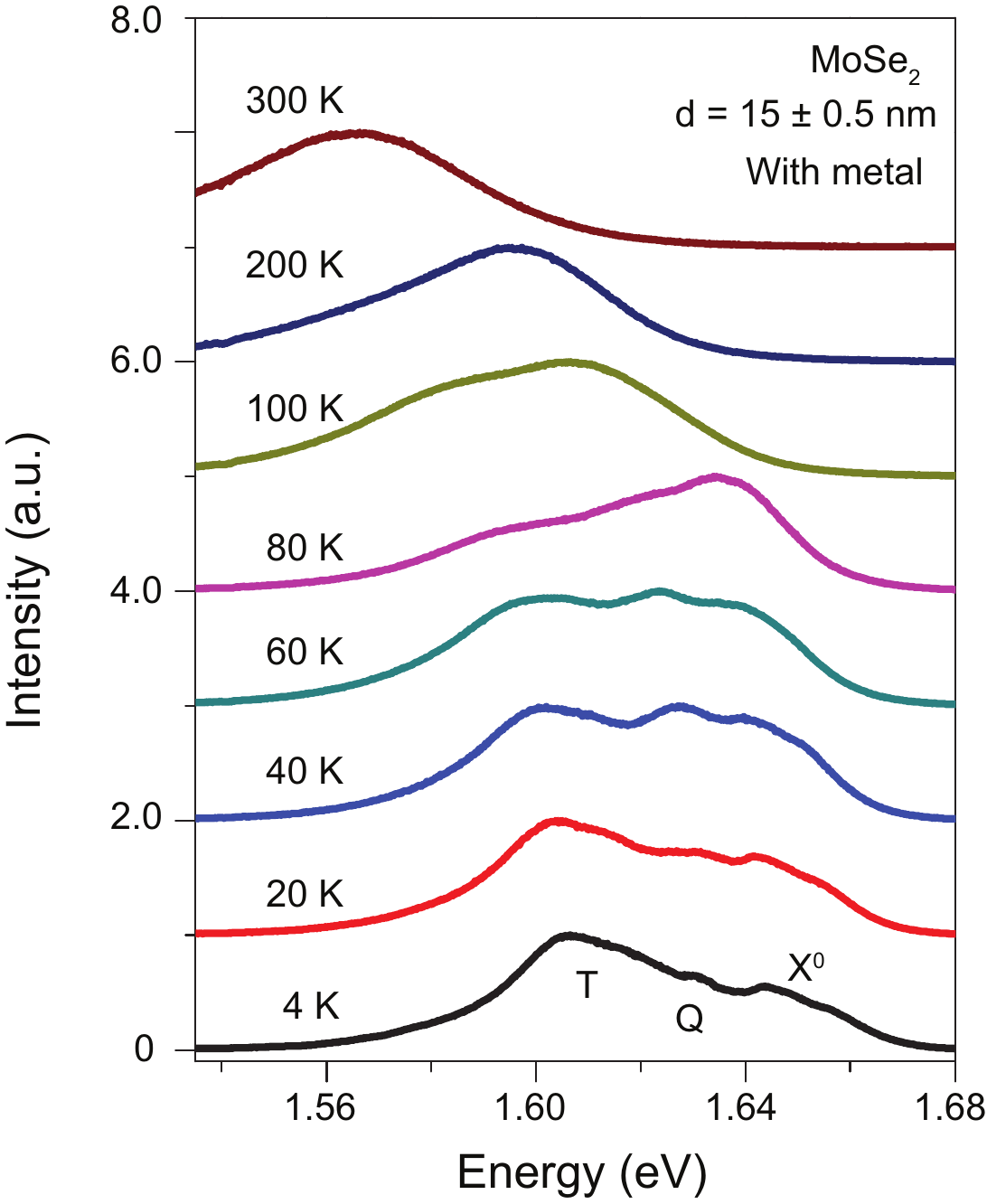}
	\end{center}
	\caption{Normalized photoluminescence spectrum for the same full structure with MoSe$_2$ used in the red curve of Figure 2(a) of the main text, for various temperatures.
	}
	\label{sifig9}
\end{figure}

\begin{figure}[h]
	\begin{center}
		\includegraphics[width = .65\linewidth]{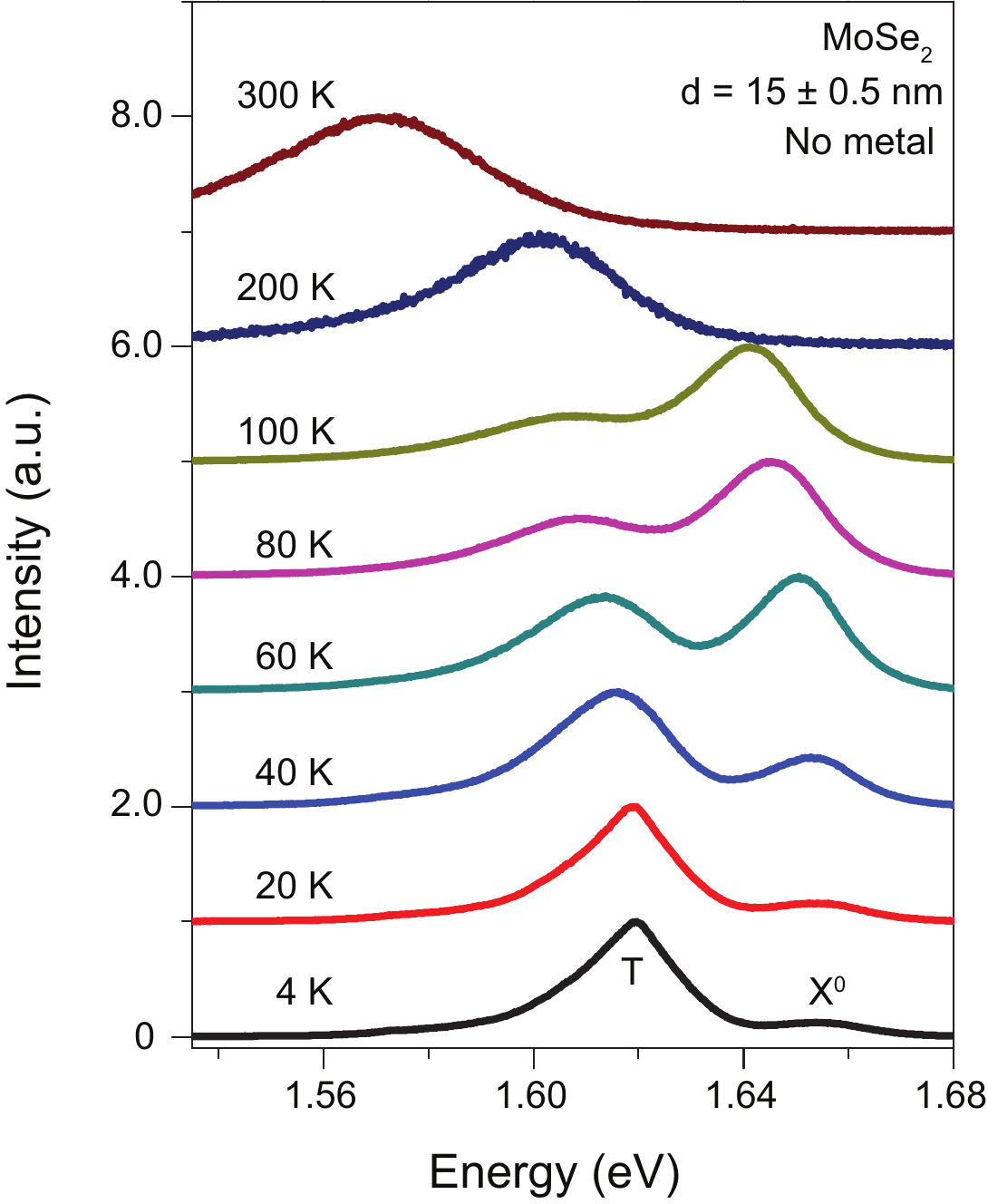}
	\end{center}
	\caption{Normalized photoluminescence spectrum for the same full structure with MoSe$_2$ used in the blue curve of Figure 2(a) of the main text, with no metal layer, for various temperatures. No Q line is observed.
	}
	\label{sifig10}
\end{figure}

\begin{figure}[h]
	\begin{center}
		\includegraphics[width = .65\linewidth]{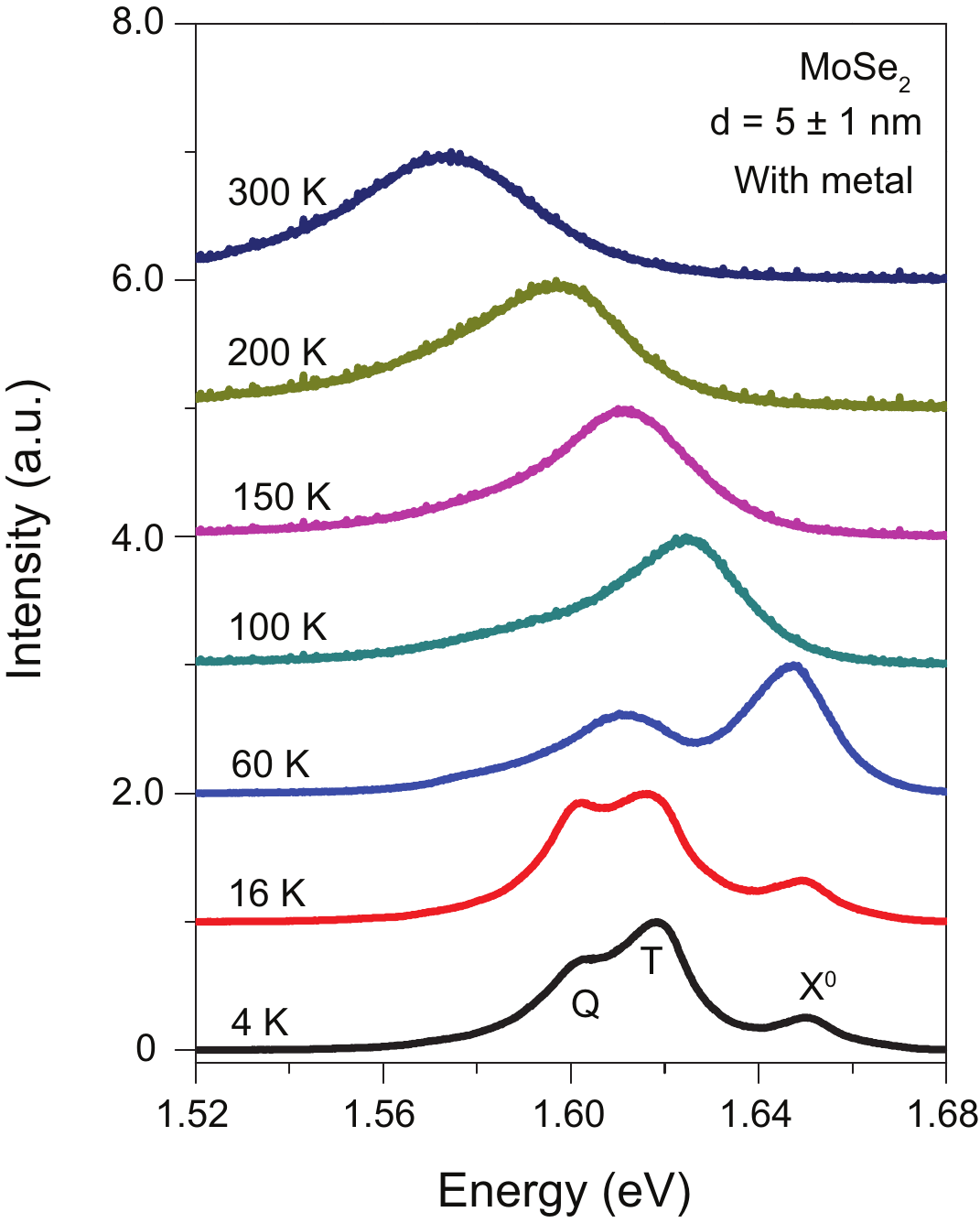}
	\end{center}
	\caption{Normalized photoluminescence spectrum for the same full structure with MoSe$_2$ used in the red curve of Figure 2(b) of the main text, for various temperatures.
	}
	\label{sifig11}
\end{figure}

\begin{figure}[h]
	\begin{center}
		\includegraphics[width = .65\linewidth]{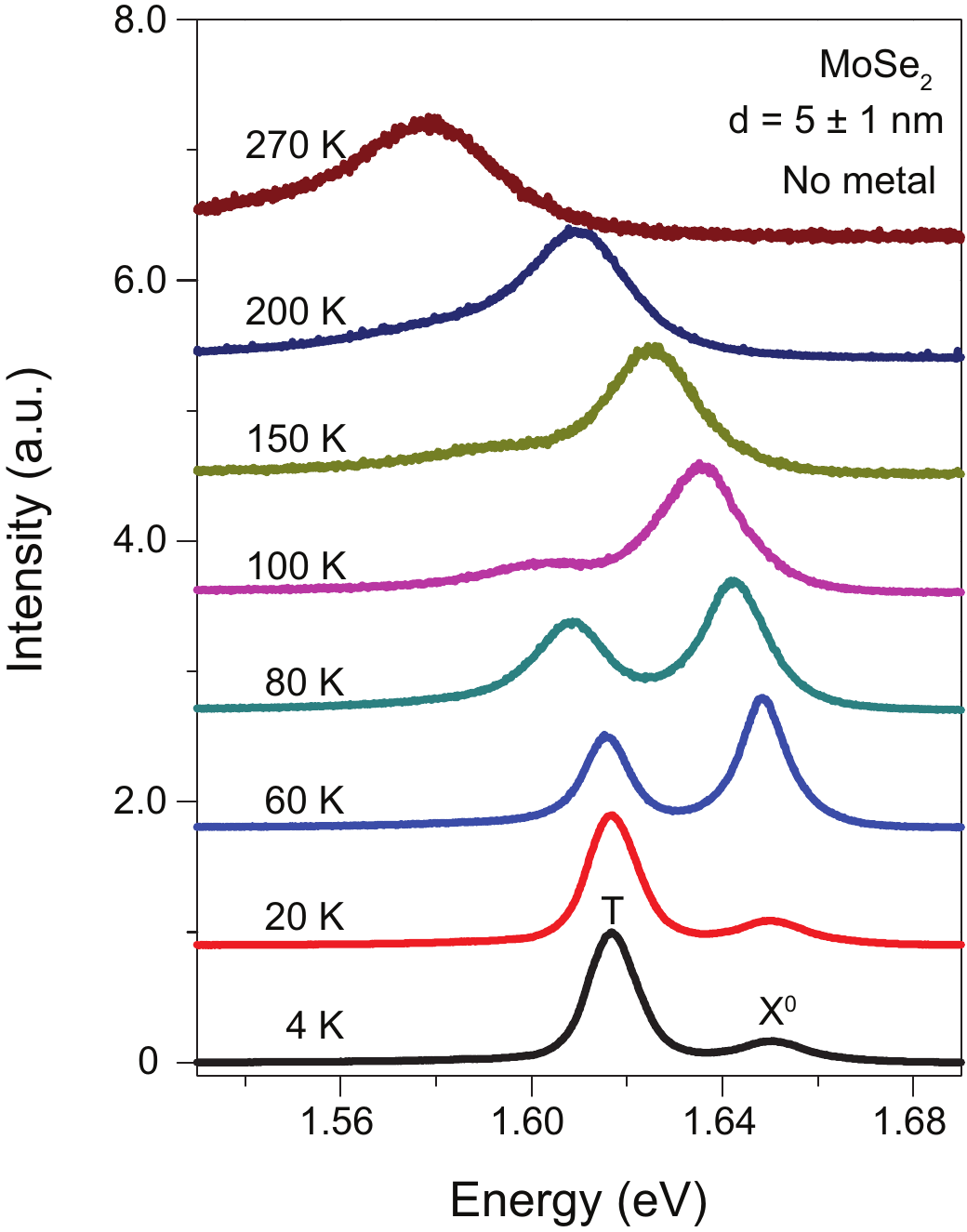}
	\end{center}
	\caption{Normalized photoluminescence spectrum for the same full structure with MoSe$_2$ used in the blue curve of Figure 2(b) of the main text, with no metal layer, for various temperatures. No Q line is observed.
	}
	\label{sifig12}
\end{figure}

{\bf Extended study of MoSe$_2$ structures}. Figures \ref{sifig9} to \ref{sifig12} show the temperature dependence of the spectra for the same two samples as used for the data of Figures 2(a) and 2(b) of the main text. 

\begin{figure}[h]
	\begin{center}
		\includegraphics[width = .85\linewidth]{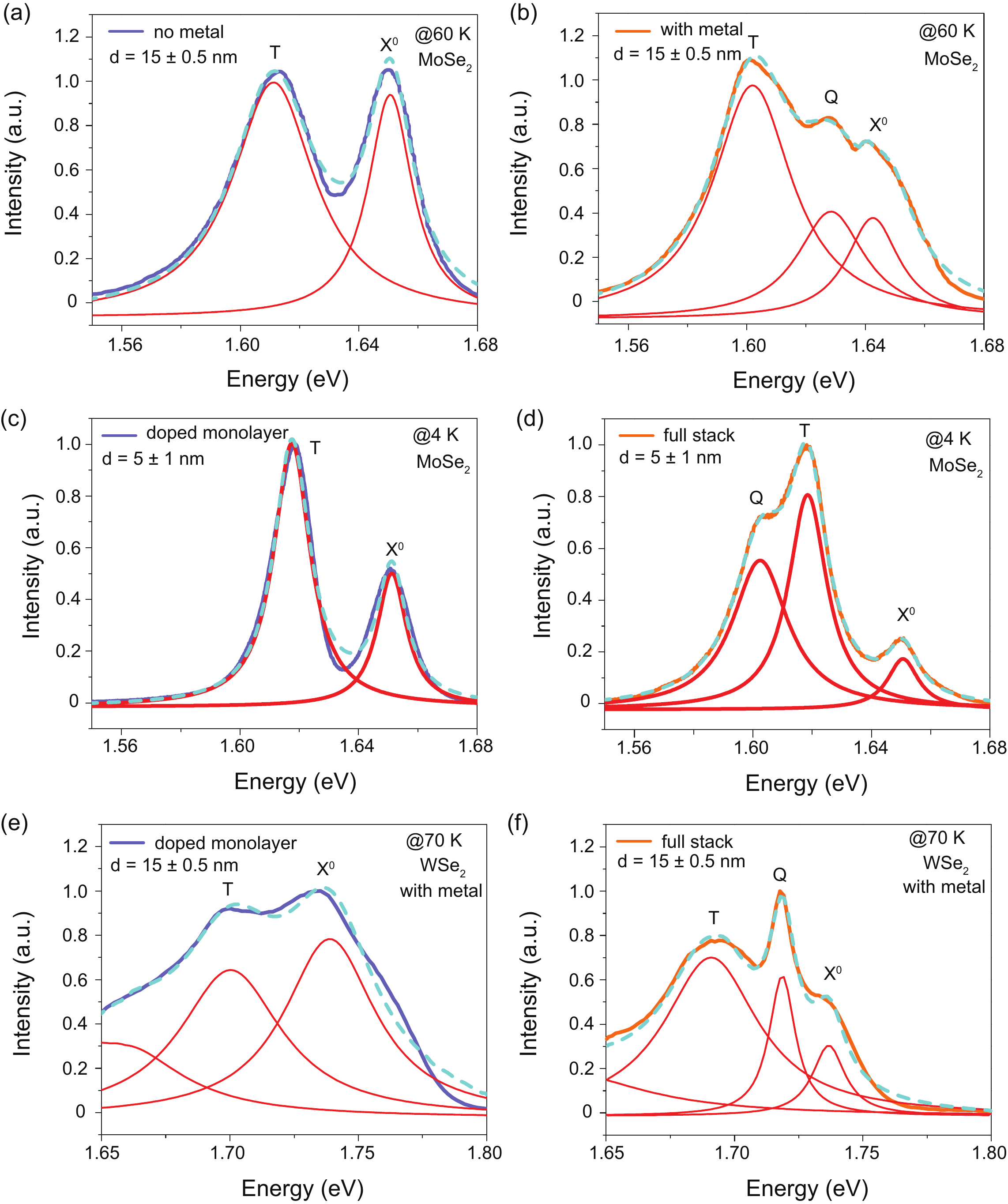}
	\end{center}
	\caption{Fits of Lorentzian peaks (solid red lines) to the data (dashed lines) from the MoSe$_2$ structures used for Figures 2(a) and 2(b) of the main text. 
	}
	\label{sifig13}
\end{figure}

Figure~\ref{sifig13} shows various fits of Lorentzian lines to the spectra of these structures at various temperatures.

\begin{figure}[h]
	\begin{center}
		\includegraphics[width = .8\linewidth]{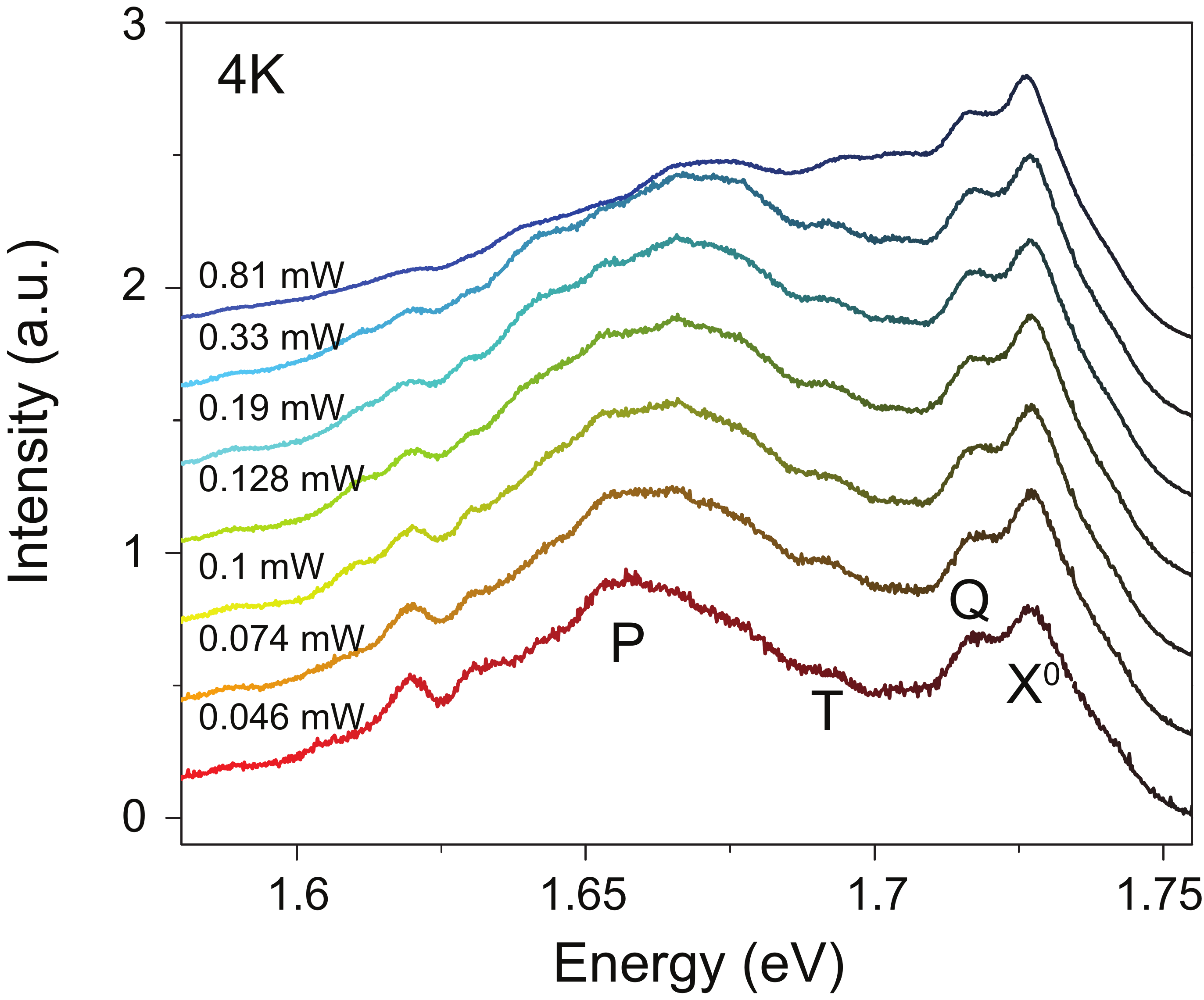}
	\end{center}
	\caption{Power dependence of a third WSe$_2$-based sample of the same design as Figures S1 and S2, pumped with a 532 nm Nd: YAG laser, at 4K. A Varshni redshift of the exciton and quaternion lines is observed due to the laser heating effect. The direct exciton is seen at 1.726 eV, and the quaternion line is also clearly identifiable around 1.714 eV. The trion line here is 1.69 eV, about 36 meV below the exciton line. A saturation of the intensity of the impurity PL is unambiguously seen, which is different from the behavior of the exciton line and the quaternion line.}
	\label{sifig6}
\end{figure}

\begin{figure}[h]
	\begin{center}
		\includegraphics[width = .98\linewidth]{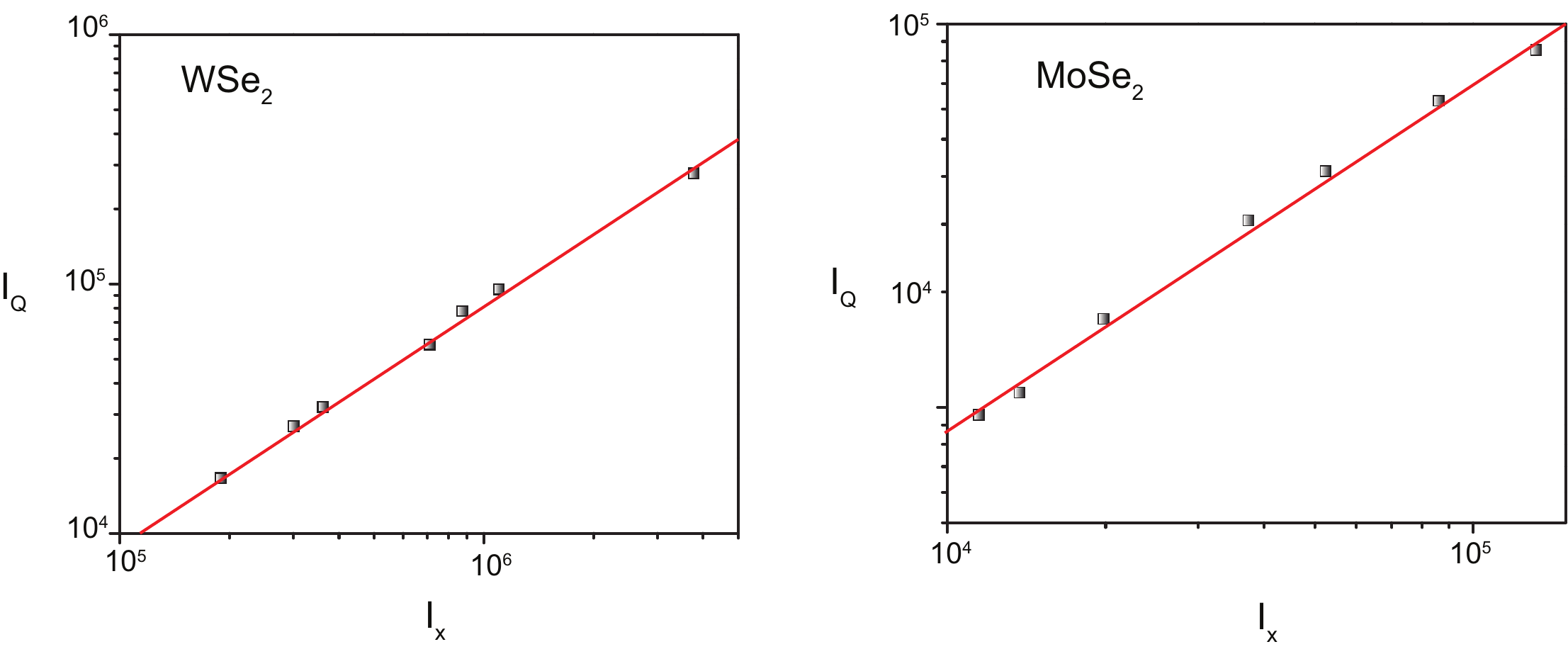}
	\end{center}
	\caption{Intensity of the PL of the Q line as a function of the exciton intensity PL for both WSe$_2$ and MoSe$_2$ stacks. The red line shows a linear dependence in both plots. The  Q line has a clear linear dependence, not superlinear, which would be expected for a biexciton complex.
	}
	\label{sifig7}
\end{figure}

{\bf Density dependence of the Q line}. 
Figure \ref{sifig6} presents a set of spectra from the full WSe$_2$ structure, showing that the Q line cannot be an impurity line, because its intensity increases as the pump power is increased, unlike impurity lines which, as expected saturate in intensity, as their number is limited. Figure \ref{sifig7} shows the intensity of the Q line relative to the exciton line intensity for both WSe$_2$ and MoSe$_2$ samples. In both cases it is clearly linear, showing that this line is not a biexciton, which is known to be proportional to the square of the exciton density, since each biexciton requires two excitons.

\newpage

\section{II. Theory for Trion Binding Energy Calculations with No Metal Present}

\begin{figure}[b]
	\includegraphics[scale=0.65]{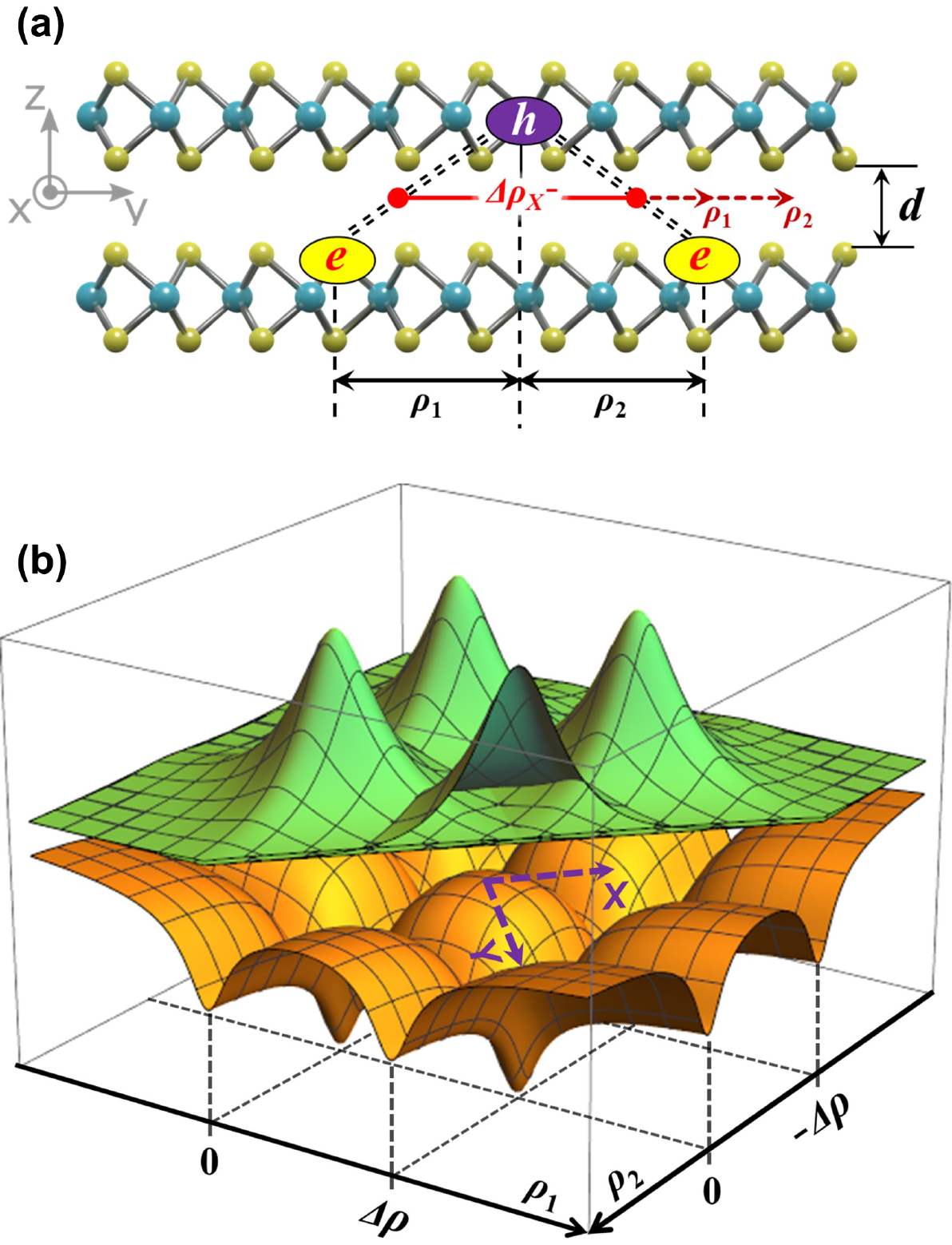}
	\caption{(a)~The structure of a negatively charged interlayer exciton (CIE) in a TMD bilayer. (b)~Schematic of the tunnel exchange coupling configuration for the two ground-state indirect excitons (IEs) to form the CIE complex in (a). The coupling occurs in the configuration space of the two \emph{independent} in-plane relative $e$-$h$ motion coordinates, $\rho_1$ and $\rho_2$, of each of the IEs separated by the center-of-mass-to-center-of-mass distance $\Delta\rho$ --- cf.~(a). The coupling is due to the tunneling of the $e$-$h$ system in $x$ (or $y$) direction through the potential barrier formed by the two $e$-$h$ electrostatic interaction potentials of the two IEs (bottom, orange color), between the equivalent IE states represented by the two-exciton wave function shown on the top (green color).}
	\label{sfig0}
\end{figure}

\subsection{General Case: Charged InterLayer Exciton in Bilayers}

We start with the most general case of the Charged Interlayer Exciton (CIE) binding energy calculation presented lately by one of us (with coauthors) in Ref.~~\onlinecite{preprint-bond}. Only a brief outline of the theory is provided here; see Refs.~~\onlinecite{BondVlad18}, ~\onlinecite{Bondarev2016}, and ~\onlinecite{preprint-bond} for particulars. A sketch of the CIE --- (interlayer) trion --- in a TMD bilayer is shown in Fig.~\ref{sfig0}(a) for the negative trion case; the positive trion can be obtained by charge sign inversion. In Figure~\ref{sfig0} and \emph{all throughout this section}, the interlayer separation distance $d$ is not to be confused with the monolayer-to-metal distance shown in Fig.~1(b) of the main text.

The CIE is a charged three-particle complex of an interlayer (indirect) exciton (IE) and an extra hole ($h$) or electron ($e$), in which two like-charge carriers confined to the same layer share an unlike-charge carrier from the other layer. Such a CIE complex can be viewed as being formed by the \emph{two} equivalent indistinguishable symmetric IE configurations with an extra charge carrier attached to the left or right IE. For such a quantum system, the effective configuration space can be represented by the two \emph{independent} in-plane projections $\rho_1$ and $\rho_2$ of the relative $e$-$h$ coordinates in each of the IEs as can be seen from the comparison of Fig.~\ref{sfig0}~(a) and (b). The CIE bound state then forms due to the exchange under-barrier tunneling in the configuration space $(\rho_1,\rho_2)$ between the two \emph{equivalent} IE configurations of the $e$-$h$ system that are separated by the IE center-of-mass-to-center-of-mass (CM) distance $\Delta\rho$. The binding strength is controlled by the exchange tunneling rate integral of the form~\cite{preprint-bond}

\begin{eqnarray}
	J_{\!X^{^{\pm}}}(\Delta\rho,\sigma,r_0,d)=2N^4\Delta\rho^2\exp\!\left[-2\alpha\!\left(\!\sqrt{\Delta\rho^2+4d^2}-2d\right)\right]\hskip3cm\label{JXdfin}\\[0.5cm]
	\times\Bigg[\frac{\alpha}{\sqrt{\Delta\rho^2+4d^2}}+\!\frac{1}{2\Big(r_0+\Big\{\!\!\begin{array}{c}1\\[-0.35cm]
			\sigma\end{array}\!\!\Big\}\Delta\rho/\lambda\Big)(\alpha\Delta\rho-1)}\Bigg]\left(\frac{r_0+\Big\{\!\!\begin{array}{c}1\\[-0.35cm]
			\sigma\end{array}\!\!\Big\}\Delta\rho/\lambda}{r_0+\Delta\rho}\right)^{\displaystyle\frac{\lambda\Delta\rho}{\Big\{\!\!\begin{array}{c}\sigma\\[-0.35cm]1\end{array}\!\!\Big\}
			\left(\alpha\Delta\rho-1\right)}}\nonumber
\end{eqnarray}
with
\begin{equation}
	\alpha=\frac{2}{1+2\sqrt{d}}~~~~\mbox{and}~~~N=\frac{4}{\sqrt{1\!+4\sqrt{d}+8d(1\!+\!\sqrt{d}\,)}}
	\label{alfaN}
\end{equation}
being the interlayer-separation dependent constants coming from the IE wave function~\cite{LeavittLittle}, and the upper or lower term is to be taken in the curly brackets for the positive or negative CIE, respectively. Here the 3D ``atomic units''\space are used~\cite{Bondarev2016,BondVlad18,preprint-bond}, with distance and energy measured in the units of the exciton Bohr radius $a^\ast_B\!\!=\!0.529\,\mbox{\AA}\,\varepsilon/\mu$ and the exciton Rydberg energy $Ry^\ast\!\!=\!\hbar^2\!/(2\mu\,m_0a_B^{\ast2})\!=\!e^2\!/(2\varepsilon a_B^\ast)\!=\!13.6\,\mbox{eV}\,\mu/\varepsilon^2$, respectively, $\varepsilon$ represents the \emph{effective} average dielectric constant of the bilayer heterostructure and $\mu\!=\!m_e/(\lambda\,m_0)$ stands for the exciton reduced effective mass (in units of free electron mass $m_0$) with $\lambda\!=\!1+m_e/m_h\!=\!1+\sigma$.

To properly take into account the screening effect for the electrostatic interaction of like charges in monolayers~\cite{KeldyshRytova,Glazov18,Crooker19}, in deriving Eq.~(\ref{JXdfin}) the Keldysh-Rytova (KR) electrostatic potential interaction energy is used in the form
\begin{equation}
	V_\texttt{KR}(\rho)=\frac{\pi}{(\epsilon_1+\epsilon_2)r_0}\left[H_{0}\!\left(\frac{\rho}{r_0}\right)-N_{0}\!\left(\frac{\rho}{r_0}\right)\right],
	\label{VKR}
\end{equation}
which we approximate in terms of elementary functions as follows (atomic units)
\[
V_{\texttt{eff}}(\rho)=\frac{1}{r_0}\left[\ln\!\left(\!1+\frac{r_0}{\rho}\!\right)+(\ln2-\gamma)e^{-\rho/r_0}\right]
\]
as was previously proposed for atomically thin layers in Ref.~\cite{Rubio11}. Here, $\rho$ is the in-plane intercharge distance, $r_0\!=\!2\pi\chi_{2D}$ is the screening length parameter with $\chi_{2D}$ being the in-plane polarizability of 2D material, and $\epsilon_{1,2}$ are the dielectric permittivities of its surroundings. For unlike charges, their screened interlayer electrostatic potential interaction energy is taken to be (atomic units)
\begin{equation}
	V_\texttt{C}(r)=-\frac{1}{r}=-\frac{1}{\sqrt{\rho^2+d^2}}
	\label{VC}
\end{equation}
as dictated by the classical electrostatic Coulomb interaction of two space-separated point charges with intercharge distance written in cylindrical coordinates.

\begin{figure}[b]
	\includegraphics[scale=0.75]{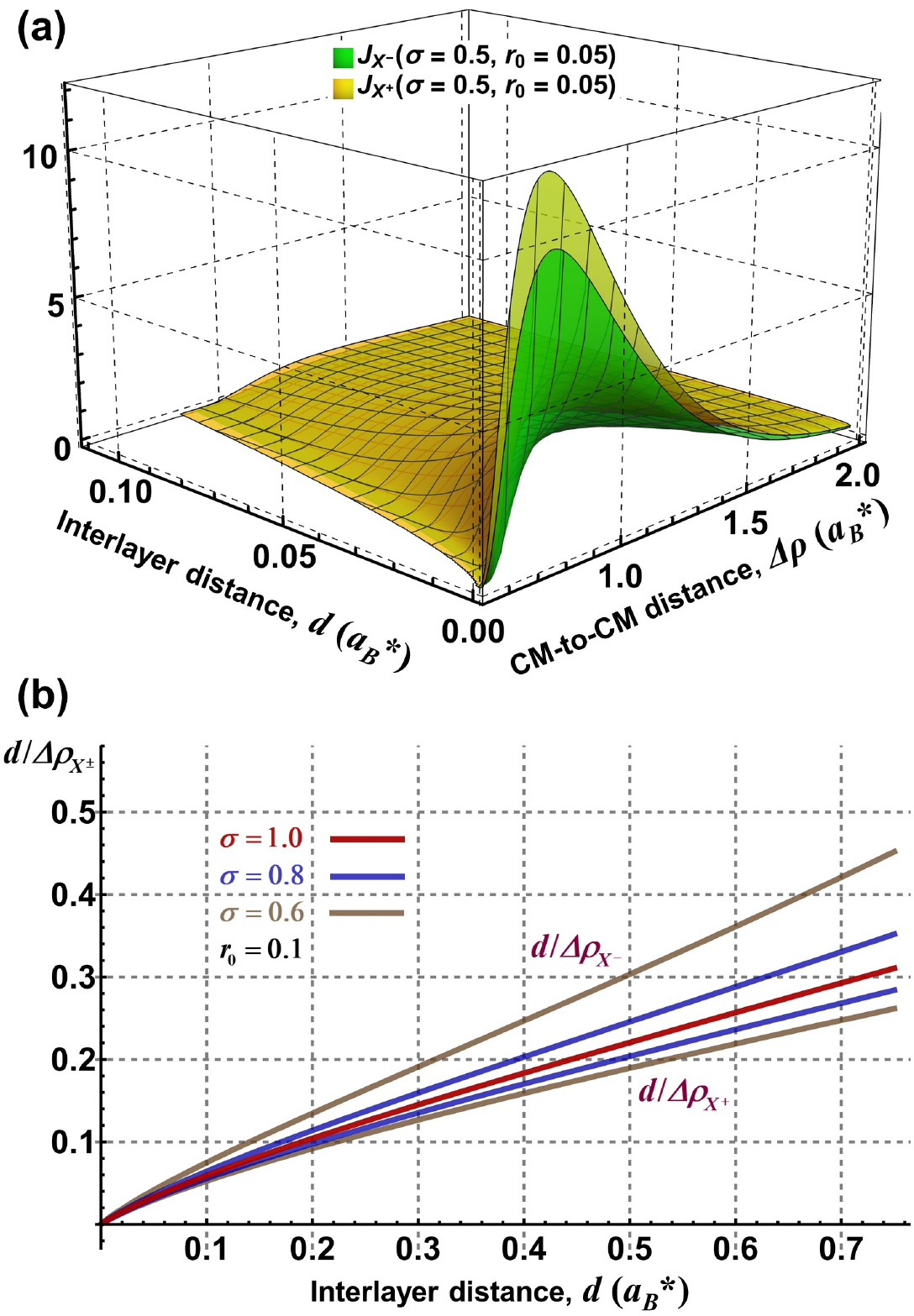}
	\caption{(a)~The exchange tunneling rate integral $J_{\!X^{^{\pm}}}$ as a function of $d$ and $\Delta\rho$, calculated per Eqs.~(\ref{JXdfin}) and (\ref{alfaN}) for a bilayer with $d$ in the range not to exceed typical van der Waals interlayer separations. (b)~The equilibrium $d/\Delta\rho_{\!X^{^{\pm}}}$ ratio as given for a similar $d$ range by Eq.~(\ref{Drho0dXpm}) with typical values of other intrinsic parameters of the bilayer system.}
	\label{sfig00}
\end{figure}

The function $J_{\!X^{^{\pm}}}$ in Eq.~(\ref{JXdfin}) is clearly seen to have a maximum as $\Delta\rho$ and $d$ vary. It tends to become a negative when $\alpha\Delta\rho<1$ in the second term in the square brackets, which is always the case for $d$ large enough whereby $\alpha\approx1/\sqrt{d}\sim0$ per Eq.~(\ref{alfaN}), making the second term in the square brackets negative and the first term negligible; whereas for $\alpha\Delta\rho>1$ it is manifestly positive and approaching zero as $\Delta\rho$ increases. As an example, in Fig.~\ref{sfig00}~(a) we show the actual behavior of the tunneling exchange integral $J_{\!X^{^{\pm}}}$ as a function of $d$ and $\Delta\rho$, calculated per Eqs.~(\ref{JXdfin}) and (\ref{alfaN}) with $d$ chosen to be in the range not to exceed typical van der Waals interlayer separations ($3-6$~\AA\space in atomic units). In the case of the CIE, $\Delta\rho>\!1$ is the physically meaningful $\Delta\rho$ domain~\cite{BondVlad18,preprint-bond}. Seeking the extremum for $J_{\!X^{^{\pm}}}(\Delta\rho)$ under this condition (with all other parameters fixed) can be done with \emph{only} the leading terms in small $1/\Delta\rho$ included in the procedure. This gives the equilibrium value of $\Delta\rho$ as follows~\cite{preprint-bond}
\begin{equation}
	\Delta\rho_{\!X^{^{\pm}}}=\frac{7\alpha-1-\Big\{\!\!\begin{array}{c}\sigma\\[-0.25cm]1/\sigma\end{array}\!\!\Big\}}{2\alpha^2}
	-\Big(3+2\Big\{\!\!\begin{array}{c}\sigma\\[-0.25cm]1/\sigma\end{array}\!\!\Big\}\Big)\,r_0\,.
	\label{Drho0dXpm}
\end{equation}
Substituting Eq.~(\ref{Drho0dXpm}) in Eq.~(\ref{JXdfin}) and reversing the sign of the result, gives the positive and negative CIE binding energies as indicated by Eq.~(1) in the main text. The equilibrium ratio $d/\Delta\rho_{\!X^{^{\pm}}}$ calculated from Eq.~(\ref{Drho0dXpm}) is shown in Fig.~\ref{sfig00}~(b).

The electrostatic interaction potential energies (\ref{VKR}) and (\ref{VC}) can be shown to consistently originate from the general solution to the electrostatic boundary-value problem that includes two coupled parallel monolayers. Such a solution was recently found and presented in Ref.~\cite{Lozovik19} (Appendix A). A bilayer system was considered to consist of the two parallel monolayers with individual $2D$-polarizabilities $\chi_\texttt{2D}^{\,\prime}$ and $\chi_\texttt{2D}^{\,\prime\prime}$ (in our notations) that are separated by a distance $d$ and surrounded by a dielectric medium of the static permittivity $\varepsilon$, with a point charge sitting at the origin of the cylindrical coordinate system placed in the bottom layer. In order to find the electrostatic interaction potential energy in the whole space, the Poisson's equation was solved in the Fourier space in the way similar to that reported in Ref.~~\onlinecite{Rubio11}. In the $2D$-coordinate space, the solution obtained yields the electrostatic unlike- and like-charge interaction energies of interest as follows (atomic units)
\begin{eqnarray}
	V_\texttt{2D}(\rho,d)=-\!\int_0^\infty\!\!\!\frac{dq\,J_0(q\rho)\,e^{-qd}}{(1+qr_0^{\prime})(1+qr_0^{\prime\prime})-q^2r_0^{\prime}r_0^{\prime\prime}\,e^{-2qd}}\,,\nonumber\\[-0.15cm]
	\label{BilayerSolved}\\[-0.15cm]
	V_\texttt{2D}(\rho,0)=\!\int_0^\infty\!\!\!\frac{dq\,J_0(q\rho)\,[1+qr_0^{\prime\prime}(1-e^{-2qd})]}{(1+qr_0^{\prime})(1+qr_0^{\prime\prime})-q^2r_0^{\prime}r_0^{\prime\prime}\,e^{-2qd}}\,,\hskip0.3cm\nonumber
\end{eqnarray}
where $r_0^\prime\!=2\pi\chi_\texttt{2D}^{\,\prime}$ and $r_0^{\,\prime\prime}\!=2\pi\chi_\texttt{2D}^{\,\prime\prime}$ are the respective screening parameters for the individual monolayers. Due to the presence of the second layer, these equations do not seem to look similar to the solitary-monolayer KR interaction case in Eq.~(\ref{VKR}) we use. However, setting $d\!=\!\infty$ to take the top layer away makes the former zero, while the latter integrates to yield the KR potential energy (\ref{VKR}) with the effective screening length $r_0=r_0^\prime$ just as it should be.

Due to the oscillatory behavior of the 0th order Bessel function $J_0(x)$ for all $x\!>\!1$, only $q\!\lesssim\!1/\rho$ contribute the most to the integrals in Eq.~(\ref{BilayerSolved}). In our case, $\rho\approx\Delta\rho_{X^\pm}$ as can be seen from Fig.~\ref{sfig0}~(a). Then, in the domain $1/\Delta\rho_{X^\pm}\!<\!1$ which Eqs.~(\ref{JXdfin}) and (\ref{Drho0dXpm}) are obtained for~\cite{BondVlad18,preprint-bond}, one has $q\!\lesssim\!1/\rho\approx\!1/\Delta\rho_{X^\pm}\!<\!1$ to contribute the most to both integrals, so that $qd\!\lesssim d/\Delta\rho_{X^\pm}\!<1$ which is indeed the case as Fig.~\ref{sfig00}~(b) shows. Therefore, it is legitimate to neglect $q^2$-terms under the integrals in Eq.~(\ref{BilayerSolved}). This gives 
\[
V_\texttt{2D}(\rho,d)\approx-\frac{1}{\rho}\int_0^\infty\!\!\frac{dx\,J_0(x)\,e^{-xd/\rho}}{1+x(r_0^\prime\!+\!r_0^{\prime\prime})/\rho}\,,~~~
V_\texttt{2D}(\rho,0)\approx\frac{1}{\rho}\int_0^\infty\!\!\!\frac{dx\,J_0(x)}{1+x(r_0^\prime\!+\!r_0^{\prime\prime})/\rho}\,,
\]
and the second integral turns into the KR potential energy (\ref{VKR}) with the screening length $r_0\!=\!r_0^{\prime}+r_0^{\prime\prime}$. Additionally, as per previous computational studies of monolayer TMDs~\cite{Berkelbach2013}, the monolayer screening length can be accurately represented by $c(\varepsilon_\perp\!-\!1)/2(\epsilon_1\!+\epsilon_2)$, where $c$ and $\varepsilon_\perp$ are the \emph{bulk} TMD out-of-plane translation period and in-plane dielectric permittivity, respectively. For a TMD bilayer embedded in hBN with $\varepsilon\!=5.87$ (averaged over all three directions~\cite{Laturia18}), which is the case for a variety of experiments~\cite{Science,Glazov18,Crooker19} including our experiment here, the typical parameters are $c\!\approx\!12\!-\!13$~\AA, $\varepsilon_\perp\!\approx\!14\!-\!17$, $\epsilon_1\!=\!(2\varepsilon_\perp\!+\varepsilon_\parallel)/3$ with $\varepsilon_\parallel\approx\varepsilon_\perp/2$~\cite{Berkelbach2013,Laturia18} and $\epsilon_2\!=\!\varepsilon$ (or vice versa), to yield $r_0\!\approx\!c(\varepsilon_\perp\!-1)/(5\varepsilon_\perp/6+\varepsilon)a_B^{\ast-1}\!<1$ as $a_B^{\ast}$ is consistently greater than $1\,\mbox{nm}$ in TMDs both by our estimates in this work and by those of the others~\cite{Berkelbach2013,Crooker19}. Then, we obtain $r_0/\rho\approx\!r_0/\Delta\rho_{X^\pm}\!\ll\!1$. With this in mind the denominator of the first integral above can be expanded in rapidly convergent binomial series, whereby after the term-by-term integration the interlayer electrostatic interaction energy takes the form
\[
V_\texttt{2D}(\rho,d)\approx-\frac{1}{\rho}\int_0^\infty\!\frac{dxJ_0(x)\,e^{-xd/\rho}}{1+xr_0/\rho}\approx-\frac{1}{\sqrt{\rho^2+d^2}}\,
\Big(1-\frac{d}{\rho}\frac{1}{1+d^2/\rho^2}\frac{r_0}{\rho}+\cdots\Big)\,.
\]
Here, the second term in parentheses comes out as the 2nd (not the 1st as one would expect!) order of smallness since $d/\rho\approx d/\Delta\rho_{X^\pm}\!<\!1$ as demonstrated in Fig.~\ref{sfig00}~(b), and so it can be safely dropped along with the rest of higher infinitesimal order terms, whereby one arrives at the interlayer Coulomb interaction (\ref{VC}) we used in our calculations throughout this work. Note also that, even more generally, this series expansion can be seen to be uniformly suitable for all $\rho\!\ge\!0$, including $\rho\!\sim\!0$ as well, in which case the second term in parentheses comes out as the 1st order of smallness in $r_0/d$ and \emph{still} can be dropped for $d$ large enough, whereby one \emph{still} arrives at Eq.~(\ref{VC}) --- now in the \emph{classical} electrostatic Coulomb interaction regime of two space-separated point charges with intercharge distance written in cylindrical coordinates, the regime we present in the main text of this work.

\subsection{Particular Case of Relevance Here: IntraLayer Trion in Monolayers}

\begin{figure}[b]
	\includegraphics[scale=0.75]{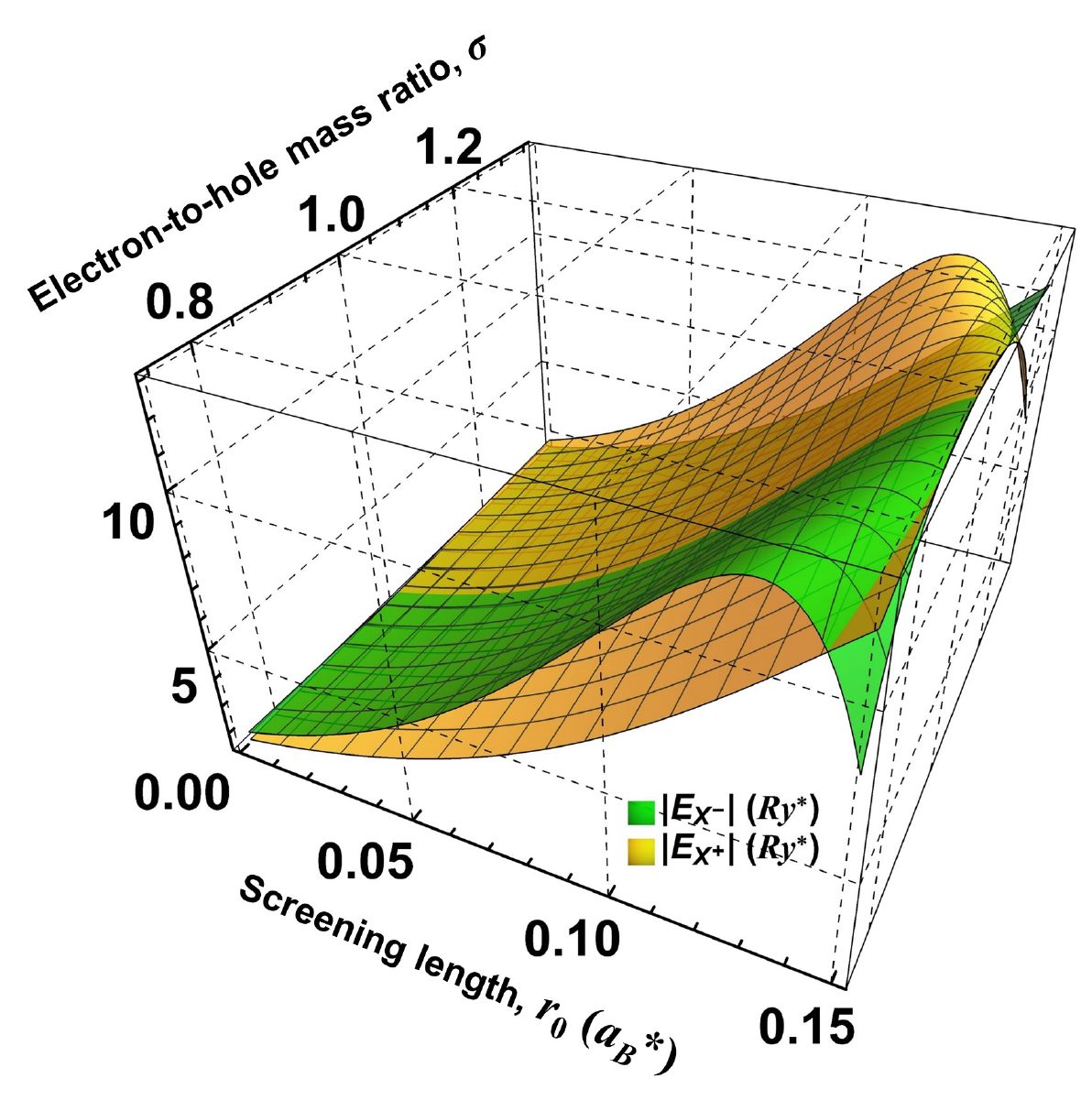}
	\caption{The absolute values of the positive and negative intralayer trion binding energies for a range of $r_0$ and $\sigma$ parameters as given by Eqs.~(\ref{JXfin}) and (\ref{Drho0Xpm}).}
	\label{sfig000}
\end{figure}

The configuration space approach outlined imposes no constraints on the interlayer spacing $d$, which can also be seen from Fig.~\ref{sfig00}~(a) where the function $J_{\!X^{^{\pm}}}$ of Eq.~(\ref{JXdfin}) remains smooth and well-defined for a large range of interlayer separation distances including $d\!=\!0$. Therefore, we use Eqs.~(\ref{JXdfin}), (\ref{alfaN}) and (\ref{Drho0dXpm}) with the $d$ parameter set equal to zero to represent a particular case of the intralayer trion we are interested in here. Plugging $d=0$ in there gives
\begin{equation}
	J_{\!X^{^{\pm}}}(\Delta\rho)\!=\!2^{10}e^{\displaystyle-4\Delta\rho}\Delta\rho
	\Bigg[1\!+\!\frac{\Delta\rho}{4\Big(r_0\!+\!\Big\{\!\begin{array}{c}1\\[-0.4cm]\sigma\end{array}\!\!\Big\}\Delta\rho/\lambda\Big)(2\Delta\rho-1)}\Bigg]\!\!
	\left(\!\!\frac{r_0\!+\!\Big\{\!\begin{array}{c}1\\[-0.4cm]\sigma\end{array}\!\!\Big\}\Delta\rho/\lambda}{r_0+\Delta\rho}\!\!\right)^{\!\!\displaystyle
		\frac{\lambda\Delta\rho}{\Big\{\!\begin{array}{c}\sigma\\[-0.35cm]1\end{array}\!\!\Big\}\!\left(2\Delta\rho-1\right)}}
	\label{JXfin}
\end{equation}
and
\begin{equation}
	\Delta\rho_{\!X^{^{\pm}}}=\frac{13-\Big\{\!\!\begin{array}{c}\sigma\\[-0.25cm]1/\sigma\end{array}\!\!\Big\}}{8}
	-\Big(3+2\Big\{\!\!\begin{array}{c}\sigma\\[-0.25cm]1/\sigma\end{array}\!\!\Big\}\Big)\,r_0\,.
	\label{Drho0Xpm}
\end{equation}
These are $J_{\!X^{^{\pm}}}$ and $\Delta\rho_{\!X^{^{\pm}}}$ referred to in the main text of our work. Plugging Eq.~(\ref{Drho0Xpm}) in Eq.~(\ref{JXfin}) with the sign reversed gives the positive $E_{X^+}$ and negative $E_{X^-}$ intralayer trion binding energies as functions of $r_0$ and $\sigma\!=\!m_e/m_h$. Their absolute values are shown in Fig.~\ref{sfig000}. For $\sigma\!\ne\!1$ a slight difference between the two can be seen to increase with increasing $r_0$, the screening length parameter. Since in our experiment $r_0$ is estimated to be small, on the order of a few hundredths within our model, we neglect this effect in our analysis presented in the main text.

\subsection{Remarks on the Near-Field Plasmonic Effects}

There are two factors one has to care about in experiments with semiconductors located near metallic surfaces. They are associated with electronic and photonic density-of-states (DOS) modifications in the near-field zone close to the metal surface.

The image-charge approach we use in our analysis is known to break down at nanoscale distances to the surface as the screening charge cannot be localized on the metal surface due to the fundamental principles of quantum mechanics~\cite{Kaiser17}. The Thomas-Fermi (TF) charge screening theory in metals takes into account the energy gain due to the screening of the external charge and the energy cost to localize the screening charge. The screening charge delocalization results in the screening at a finite wavelength $l_{TF}$, called the TF screening length, controlled by the \emph{electronic} DOS at the Fermi level. The TF theory was recently extended to show a profound strength reduction for the electrostatic Coulomb interaction at distances under $l_{TF}$, with no change to the image-charge theory predictions otherwise~\cite{Kaiser17}. Estimates of typical $l_{TF}$ values for metals can be obtained from the ratio of their plasma frequency $\omega_p$ to Fermi velocity $v_F$, that is $l_{TF}\!=\!(\omega_p/v_{F})^{-1}\!\lesssim\!1$~\AA. For example, for Nb used in our experiments we have $\omega_p\!=\!8.87$~eV and $v_F\!=\!0.61\!\times10^8$~cm/s~\cite{Chakrab76}, to obtain $l_{TF}\!\approx\!0.44$~\AA. This is definitely a negligible quantity as compared to $d = 5$ to 15~nm between the metal surface and closest TMD monolayer in our samples, suggesting that the image-charge analysis is fully legitimate in our case. Moreover, as typical van der Waals distances are known to be in the range of 3 to 6~\AA, even a TMD monolayer placed directly on the metal surface could still be described reasonably well in terms of the electrostatic image-charge approach.

Another factor to take into consideration is local \emph{photonic} DOS variations associated with the plasma excitations on the metal surface due to the exciton-plasmon coupling. For a dipole emitter, which is the intralayer exciton in our case, this could potentially red-shift or even split the PL emission lines, also redistributing their intensities in our experiments as one gets closer to the metal surface by decreasing the TMD monolayer-to-surface distance. Even the metal layer thickness matters in this case as per the most recent theoretical analysis~\cite{BondShal20}, however, not for macroscopically thick metallic substrates with the dipole-emitter-to-surface distances we have in our experimental samples, so that the exciton-plasmon coupling effects can still be safely neglected.
